\documentclass[conference]{IEEEtran}
\IEEEoverridecommandlockouts
\usepackage{cite}
\usepackage{amsmath,amssymb,amsfonts}
\usepackage{algorithmic}
\usepackage{graphicx}
\usepackage{textcomp}
\usepackage{xcolor}
\def\BibTeX{{\rm B\kern-.05em{\sc i\kern-.025em b}\kern-.08em
    T\kern-.1667em\lower.7ex\hbox{E}\kern-.125emX}}
\usepackage{cuted}
\usepackage{caption}
\usepackage{marvosym}
\usepackage{float}
\usepackage[final]{changes}

\begin{document}

\title{SeriCrypt: An LLM-Driven Context-Aware Serialization Framework for Cryptographic Protocols
}
\author{\IEEEauthorblockN{Maosong Chen$^{\dagger}$, Xi Chen$^{\dagger}$\textsuperscript{\Letter}, Mengcheng Ju$^{\dagger}$, Dongliang Zhao$^{\dagger}$, Chunxiang Gu$^{\dagger}$\textsuperscript{\Letter}}
\IEEEauthorblockA{$^{\dagger}$Information Engineering University, Zhengzhou 450001, China \\
Email: chenmaosong1214@163.com, xycuckoo@tsinghua.org.cn, zhangph12138@163.com, xiao\_\_xiang@163.com, gcx5209@126.com}
}
\maketitle

\begin{abstract}
Constructing syntactically correct and cryptographically valid message sequences is essential for protocol state machine learning, conformance testing, and fuzzing. Unlike plaintext protocols, cryptographic protocols involve complex cross-message state dependencies and cryptographic computation constraints. Existing automated message construction approaches predominantly target text-based or plaintext protocols, lacking support for deep semantic dependencies and cryptographic operations, rendering cryptographic protocol message construction largely manual. We present SeriCrypt, an LLM-driven, context-aware serialization framework for cryptographic protocols. SeriCrypt uses a large language model (LLM) to extract protocol field constraints, contextual state dependencies, and cryptographic computation rules from unstructured protocol specifications, representing them as a unified structured intermediate representation. To formally characterize this representation, we design a domain-specific language for cryptographic protocols (CDSL) that explicitly captures semantic constraints and computational dependencies in protocol interactions. Building upon CDSL, we implement a protocol-agnostic execution engine that parses CDSL declarations and automates field value resolution, cryptographic primitive invocation, and byte-stream serialization. \replaced{As case studies of SeriCrypt in protocol security testing,}{The general serialization capability provided by SeriCrypt supports various protocol security testing tasks. As a case study,} we employ the framework to construct violation messages targeting specification-defined security constraints \replaced{and validate its application in protocol fuzzing scenarios, with evaluation on}{and evaluate them on} multiple mainstream implementations of \replaced{TLS 1.2/1.3, IKEv1/v2, SSH, and TLCP}{TLS 1.2/1.3 and IKEv1/v2}. Results show SeriCrypt can generate valid message sequences accepted by real implementations and successfully complete handshakes across all evaluated scenarios. Security constraint testing revealed five specification violations, \replaced{and fuzzing reached deeper protocol states with higher code coverage than mainstream fuzzers under the same time budget, demonstrating the framework's practical value for cryptographic protocol security testing.}{demonstrating the framework's practical value for cryptographic protocol security testing.}
\end{abstract}


\section{Introduction}

Cryptographic network protocols (e.g., TLS, IPsec, SSH) are the foundational infrastructure of modern network communication security, providing core functions such as confidentiality protection, identity authentication, and integrity verification. The security of these protocols depends not only on the completeness of the specification design but also on whether their implementations strictly adhere to the security constraints defined in the RFCs. Studies show that even when the specification itself is thoroughly verified, deviations at the implementation level can still lead to severe security vulnerabilities such as state machine bypass \cite{fiterau2020analysis,fiterau2023automata,baumer2025finding,baumer2024terrapin}. Therefore, security testing against protocol implementations is an indispensable means of ensuring communication security \cite{zhao2025aglfuzz}.

For encrypted protocols, constructing messages that conform to cryptographic specifications can effectively improve the efficiency of security testing \added{methods such as fuzzing~\cite{ammann2024dy}, state-machine learning~\cite{fiterau2023automata}, and conformance testing}. This requires that messages not only satisfy precise byte-level encoding but also handle complex cryptographic computations within a complete session context, and that their values hold deep dependencies on the states of preceding interactions.

This strict requirement of cross-message state consistency and cryptographic correctness poses \replaced{challenges}{a dual challenge} \replaced{for automating the construction of such messages}{for automated protocol testing}. \added{The knowledge required to construct such messages is described in natural language in protocol specifications (e.g., IETF RFCs and national standards). LLMs have natural language understanding capabilities and can automatically extract this protocol knowledge from the specification text, but two problems remain in their practical application:} First, directly employing LLMs to generate binary messages or low-level handling code often fails to reliably meet the above constraints \cite{li2025cipherbank}. Second, although existing automation schemes have attempted to introduce formal protocol description languages to constrain LLMs in automatically extracting message layout, length, and nesting logic \cite{fakhoury20253dgen,zheng2025validating}, their research focus has primarily been on parsing and validating message syntactic structure, without covering cryptographic semantics such as key derivation, encryption encapsulation, and signature verification. As a result, even if developers can automatically obtain the syntactic skeleton and static attributes of messages, they still need to manually write substantial state maintenance and cryptographic computation code to construct valid messages that are actually accepted by the server.

To address this problem, this paper proposes SeriCrypt, a context-aware serialization framework for cryptographic protocols. The core idea of SeriCrypt is to decouple protocol structure description from cryptographic attribute computation. The LLM is only responsible for extracting a structured protocol intermediate representation—the Protocol Abstract Syntax Tree (PAST)—that conforms to the constraints of CDSL. PAST organizes message formats in a hierarchical structure, and defines field lengths, cross-message references, key derivation, and encryption/decryption processes through attribute rules. Then the execution engine parses these attribute rules, calls underlying cryptographic primitives, and completes the evaluation and serialization. \replaced{This design automates the construction of cryptographic protocol messages, and further allows testers to modify message structures at the PAST level to generate test sequences that conform to cryptographic specifications, thereby supporting security testing of protocol implementations.}{This design enables testers to modify message structures at the PAST level and generate test messages that conform to cryptographic specifications, thereby supporting security testing of protocol implementations.}

We have implemented SeriCrypt and evaluated it on mainstream open-source implementations including OpenSSL, TLSE, strongSwan, \replaced{Libreswan, OpenSSH, and GmSSL}{and Libreswan}. The experiments cover \replaced{six cryptographic protocols: TLS 1.2/1.3, IKEv1/v2, SSH, and TLCP}{four protocol versions: TLS 1.2/1.3 and IKEv1/v2}. The results show that SeriCrypt can automatically construct acceptable legitimate message sequences and successfully complete full handshakes. In addition, \replaced{we applied SeriCrypt to protocol security testing scenarios and discovered five security violations.}{we constructed targeted violation messages against RFC security constraints for testing and discovered five security violations against the specifications.}

The contributions of this paper include:
\begin{itemize}
\item \textbf{Context-aware protocol representation.} We designed a domain-specific language for cryptographic protocols, CDSL, for uniformly describing message structures, field attributes, cross-message context references, and cryptographic computation rules such as key derivation, encryption, and authentication.
\item \textbf{Specification-driven automatic extraction process.} We propose an LLM-based specification extraction chain-of-thought that extracts the Protocol Abstract Syntax Tree based on the LLM's existing protocol knowledge, reducing the cost of manually organizing protocol messages and cryptographic dependency rules.
\item \textbf{Protocol-independent general execution engine.} We implemented a modular execution engine for CDSL attributes that completes field evaluation, cryptographic operation execution, and message serialization, so that protocol differences are mainly reflected in the CDSL description layer while the underlying execution mechanism can be reused across protocols.
\item \textbf{Experimental validation and security testing application.} Through end-to-end evaluation of \replaced{six mainstream implementations across six cryptographic protocols}{four mainstream implementations across four protocol versions}, \replaced{we verified the interoperability of the messages generated by SeriCrypt for cryptographic protocols}{we verified the generation correctness of SeriCrypt for cryptographic protocols}. \replaced{Meanwhile, experiments demonstrate its supporting capability in protocol security testing scenarios, with five security violations discovered.}{Meanwhile, five security violations were discovered through targeted violation testing, demonstrating the framework's supporting capability for protocol security testing.}
\end{itemize}

\section{Problem Definition and Design Requirements}

The generation of cryptographic protocol messages is fundamentally different from that of plaintext protocols. For cryptographic protocols, even if the precise syntactic structure of protocol messages is known, if one cannot access and process the contextual information and cryptographic computations in the session process, it is still impossible to construct a message that can pass peer verification \cite{bhargavan2009cryptographic,dong2012cryptographic}. The complexity is mainly reflected in two aspects: cross-message context references---some fields need to reference random numbers, negotiated parameters, and ephemeral keys exchanged or negotiated in previous messages; cryptographic computation constraints---some fields are themselves the results of cryptographic operations, such as key derivation, message authentication codes, and digital signatures.

Take the Finished message in TLS~1.3 as an example \cite{rescorla2018transport}. The syntactic structure of this message is simple, but the construction of its verify\_data field must take as input the transcript-hash accumulated from previous handshake messages, combine it with the finished-key derived from the shared secret, and then compute the HMAC. Relying solely on static field format definitions cannot satisfy such dynamic computation requirements.

Accordingly, this paper decomposes the problem of automated construction of cryptographic protocol messages into two sub-problems:
\begin{itemize}
\item\textbf {Representation problem} --- a formal language is needed to describe protocol message structures, cross-message state dependencies, and cryptographic computation logic, in order to automatically construct the protocol's intermediate representation.
\item\textbf {Execution problem} --- a general execution engine decoupled from specific protocols is needed, which can parse the aforementioned intermediate representation and drive the underlying cryptographic primitives to complete evaluation and serialization.
\end{itemize}

The system design needs to meet the following requirements:
\begin{itemize}
\item \textbf{R1} Structural expressiveness: The formal language must fully describe the organization of protocol messages, including field order, hierarchical relationships, length constraints, optional fields, and nested structures.
\item \textbf{R2} Dependency expressiveness: The formal language must explicitly express the dependence of field values on the results of preceding interactions, making the sources of cross-message references traceable.
\item \textbf{R3} Cryptographic computation expressiveness: The formal language must explicitly express the cryptographic computations involved in field assignments, concretizing the computation methods of relevant fields.
\item \textbf{R4} Executability: The formal language must adopt a uniform, unambiguous formal definition, ensuring that the intermediate representation constructed under its constraints is machine-readable and can serve as input to the execution engine.
\item \textbf{R5} Attribute-aware general parsing: The execution engine must have parsing capability independent of specific protocols, driving evaluation and serialization based on the attributes of nodes in the intermediate representation.
\end{itemize}

\section{Formal Model and Unified Attribute System}

This section establishes the formal foundation of SeriCrypt. We propose the Context-Aware Attribute Grammar for Cryptographic Protocols (CA-AGCP) model and, on this basis, define the Unified Attribute System (UAS).

\subsection{From CFG+AG to CA-AGCP}

In classical formal language theory \cite{hopcroft2001introduction}, a context-free grammar (CFG) is represented as a quadruple $G = (N, \Sigma, P, S)$. CFGs can naturally describe the hierarchical nested structure of protocol messages through recursive structures. Attribute grammars (AG) \cite{knuth1968semantics} introduce attribute sets $A$ and semantic rules $R$ on this basis, and achieve node evaluation through synthesized and inherited attributes. However, applying classical AGs to cryptographic protocols requires solving two additional modeling challenges: first, dynamic awareness of cross-message states --- the field values of subsequent messages depend on the exchange results of previous messages, which extend beyond the local attribute propagation scope of standard AGs; second, primitive representation of cryptographic operators --- key derivation, encryption/decryption encapsulation, and signature generation are usually composed of a series of cryptographic primitives, and the semantic rules of standard AGs are difficult to formally characterize such operations.\added{These two challenges shape the extension design of CA-AGCP: a runtime context $\Gamma$ is introduced to maintain the cross-message state exchanged during the session, a cryptographic primitive space $\Omega$ to characterize cryptographic operations as ordered compositions of atomic primitives, complemented by a knowledge space $K$ supplying the protocol constants.}

\subsection{Formal Definition of CA-AGCP}
\deleted{CA-AGCP solves the above problems by integrating dynamic context and cryptographic primitive space into the traditional model.} CA-AGCP is formally defined as a 9-tuple $G = (N, \Sigma, P, S, A, R, $K$, \Gamma, \Omega)$, which can conceptually be divided into three layers: $\langle N, \Sigma, P, S \rangle$ forms the syntactic skeleton, defining the hierarchical basic topological structure of the protocol; $\langle A, R \rangle$ forms the basic attribute system, providing a node-level attribute evaluation framework; $\langle K, \Gamma, \Omega \rangle$ forms the dynamic semantic extension, which is the core increment of CA-AGCP relative to traditional attribute grammars. 

\textbf{Definition 3.1 (Configuration and Knowledge Space).} The knowledge space K is defined as a finite total function from identifiers to constant values:
\[K: \text{ConstId} \to \text{Val}_k\]

where ConstId is the set of constant identifiers defined by the protocol specification (e.g., algorithm identifiers, version numbers, fixed label strings, etc.), and Val covers data types such as byte sequences and certificates. K is jointly constituted by the protocol specification, local security policies, and the current test configuration.

\textbf{Definition 3.2 (Runtime Context).} The runtime context \(\Gamma\) is defined as a finite partial function from structured identifier paths to a value domain:
\[\Gamma: \text{KeySpace} \to \text{Val}\]

The key space contains two addressing modes: structured paths (used to locate field data in exchanged messages) and simple identifiers (used to name intermediate results produced during key derivation). \(\Gamma\) supports a projection operation (read) and an update operation (write). The read operation projects the specified value from the context, and the write operation injects newly generated field values or intermediate keys into the session state.

\textbf{Definition 3.3 (Cryptographic Primitive Space).} The cryptographic primitive space \(\Omega\) is defined as a set of cryptographic operations with a two-layer structure:
\[\Omega = \Omega_{\text{prim}} \cup \Omega_{\text{comp}}\]

where \(\Omega_{\text{prim}}\) is the set of atomic primitives, each element directly mapping to a single algorithm call of the underlying cryptographic library (e.g., HMAC, AES-GCM-Encrypt, etc.); \(\Omega_{\text{comp}}\) is the set of composite operations, consisting of ordered rule chains of atomic primitives and auxiliary functions according to data dependencies (e.g., key derivation, encryption/decryption encapsulation, etc.).

\textbf{Definition 3.4 (Protocol Abstract Syntax Tree).} Given a CA-AGCP grammar G, PAST is defined as a 2-tuple:
\[T_A = (T, B)\]

where T is the message structure tree, with each node carrying attribute annotations; B is the set of session-level computation blocks, each block consisting of an ordered rule chain of atomic primitives, auxiliary functions, and intermediate variables, used to generate key variables or define composite cryptographic operations. T and B are associated through the dynamic context \(\Gamma\).

\subsection{Unified Attribute System (UAS)}

To characterize the context dependencies and cryptographic transformations of cryptographic protocols, we construct the UAS, which divides the attribute set $A$ into the following three categories according to semantic function and computation method:

\textbf{Intrinsic Attributes ($A_I$):} Describe context-free atomic data units. Their attribute values are determined by the static semantics of the symbols themselves or local encoding rules, and do not depend on \(\Gamma\) or \(\Omega\). For an arbitrary symbol \( X \), with its attribute \( a \in A_I(X) \), the evaluation rule is formally expressed as:
\[\text{val}(a) = \Phi(X, K)\]

where \(\Phi: (N \cup \Sigma) \times K \to \text{Val}\) is the independent evaluation function.

\textbf{Relational Attributes ($A_R$):} Represent data dependencies between protocol fields formed based on historical interactions. Their attribute values are highly dependent on the dynamic context \(\Gamma\). For an arbitrary symbol \( X \), with its attribute \( a \in A_R(X) \), the evaluation rule is formally expressed as:
\[\text{val}(a) = \pi(\Gamma, \text{path}(a))\]

where \(\pi\) is the state projection function, and \(\text{path}(a) \in \text{KeySpace}\) is the context reference path declared by attribute \( a \).

\textbf{Transformational Attributes ($A_T$):} Characterize the higher-order semantics driven by cryptographic primitives in the protocol, such as ciphertext generation, hash verification, and key derivation. For an arbitrary symbol \( X \), with its attribute \( a \in A_T(X) \), the evaluation rule is formally expressed as:
\[\text{val}(a) = \omega(\text{inputs}(a, \Gamma, K))\]

where \(\omega \in \Omega\) is a specific cryptographic operation, and \(\text{inputs}(a, \Gamma, K)\) is the tuple of input parameters extracted from the context \(\Gamma\) and the knowledge space \(K\) according to the declaration of attribute \( a \).

\section{CDSL: Domain-Specific Language for Cryptographic Protocols}

CA-AGCP provides a formal foundation for the structural representation and semantic evaluation of cryptographic protocol messages. This section instantiates the model into an unambiguous, machine-parsable formal language --- CDSL. The problem CDSL needs to solve is: how to simultaneously carry both the structural dimension and the semantic dimension from the 9-tuple within a unified grammatical framework. To this end, CDSL designs three subsystems: Message Structure Grammar, Unified Attribute Evaluation System, and Session-Level Cryptographic Computation Grammar. Examples of PAST in subsequent subsections can be found in Appendix~A. \deleted{Table~\ref{tab:cdsl-mapping} presents the correspondence between the CDSL subsystems and the elements of the CA-AGCP 9-tuple.}


\subsection{Message Structure Grammar}

The structural grammar of CDSL instantiates the syntactic framework $\langle N, \Sigma, P, S \rangle$ of CA-AGCP as a set of named production rules. The following BNF defines its abstract syntax:

\begin{align*}
\langle \textit{Spec} \rangle &::= \text{`\{'}\ \langle \textit{RuleList} \rangle\ \text{`\}'} \\
\langle \textit{RuleList} \rangle &::= \langle \textit{Rule} \rangle \mid \langle \textit{Rule} \rangle\ \text{`,'}\  \langle \textit{RuleList} \rangle \\
\langle \textit{Rule} \rangle &::= \textit{RuleName}\ \text{`:'}\  \langle \textit{Alternatives} \rangle \\
\langle \textit{Alternatives} \rangle &::= \langle \textit{Branch} \rangle \mid \langle \textit{Branch} \rangle\ \text{`\textbar'}\  \langle \textit{Alternatives} \rangle \\
\langle \textit{Branch} \rangle &::= \langle \textit{RefSeq} \rangle \mid \langle \textit{AtomicNode} \rangle \\
\langle \textit{RefSeq} \rangle &::= \langle \textit{RefNode} \rangle \mid \langle \textit{RefNode} \rangle\  \langle \textit{RefSeq} \rangle \\
\langle \textit{RefNode} \rangle &::= \textit{`"name"'}\ \text{`:'}\  \textit{RuleName} \\
\langle \textit{AtomicNode} \rangle &::= \textit{`"field\_length"'}\ \text{`:'}\  \langle \textit{LenSpec} \rangle\ \text{`,'} \\
&\quad\quad \text{`"value"'}\ \text{`:'}\  \langle \textit{AttrSpec} \rangle \\
\langle \textit{LenSpec} \rangle &::= \textit{length}\ \text{`:'}\  \text{`"'}\ \textit{INT}\ \textit{`bits'}\ \mid\ \textit{variable} \\
\langle \textit{AttrType} \rangle &::= \textit{static} \mid \textit{random} \mid \textit{local} \mid \textit{dependent} \\
&\quad\quad \mid\ \textit{length} \mid \textit{calculate}
\end{align*}

The design of this grammar is derived from a systematic induction of cryptographic protocol specifications such as IKEv1/v2 and TLS~1.2/1.3 \cite{rescorla2018transport,harkins1998rfc2409,kaufman2014internet,dierks2008transport}. Specifically, $\langle \textit{AtomicNode} \rangle$ maps to a terminal symbol $\sigma \in \Sigma$, representing the smallest indivisible data unit, where each instance carries both a length constraint ($\langle \textit{LenSpec} \rangle$) and an attribute annotation ($\langle \textit{AttrSpec} \rangle$); $\langle \textit{RefSeq} \rangle$ is composed of one or more $\langle \textit{RefNode} \rangle$ arranged in sequence, with each reference node pointing to the production of another rule via \textit{RuleName}, expressed as $X \to Y_1, Y_2, \ldots, Y_k$, to represent the ordered concatenation of multiple substructures; $\langle \textit{Alternatives} \rangle$ models variable-length homogeneous sequences, corresponding to the recursive production $X \to YX \mid \epsilon$, whose termination condition is governed by local configuration.

\subsection{Unified Attribute Evaluation System}

At the theoretical level, UAS classifies field attributes into three categories: $A_I$, $A_R$, $A_T$. CDSL instantiates this classification into six concrete attribute types, each carrying deterministic evaluation semantics.


Intrinsic Attributes $A_I$ have the syntax form:
\begin{align*}
\langle \textit{IntrinsicAttr} \rangle ::=\ & \textit{static(} \textit{Value} \textit{)} \mid \textit{random(} \text{`"'}\ \textit{INT}\ \textit{bits} \texttt{)} \\
\mid\ & \textit{local(} \textit{ConfigKey} \textit{)}
\end{align*}

Among them, \textit{static} binds a field to a definite value known from the protocol specification\added{, resolved by the LLM based on the specification and the session configuration}; \textit{random} is generated by a cryptographically secure pseudo-random number generator according to the field length; \textit{local} imports certificates, timestamps, etc., from local configuration.

Relational Attributes $A_R$ have the syntax form:
\begin{align*}
\langle \textit{RelationalAttr} \rangle &::= \textit{dependent(} \textit{RefPath} \textit{)} \mid \textit{SimpleId} \\
\langle \textit{RefPath} \rangle &::= \textit{Ident}\ \texttt{(} \textit{Cat} \texttt{)}\ \textit{[} \textit{ByteRange} \textit{]}
\end{align*}

Among them, \textit{dependent} references a specific field in previous messages or session state; \textit{length} is an instance of synthesized attribute, whose value is obtained from the total number of bytes after serializing a specified subtree, backfilled by the execution engine after the evaluation of child nodes is completed.

Transformational Attributes $A_T$ have the syntax form:
\begin{align*}
\langle \textit{CalcExpr} \rangle &::= \textit{calculate(} \langle \textit{PostLayer} \rangle \textit{)} \\
\langle \textit{PostLayer} \rangle &::= \langle \textit{CryptoLayer} \rangle \\
&\quad \mid f_{\text{aux}} \texttt{(} \langle \textit{CryptoLayer} \rangle \texttt{)} (\texttt{[} \langle \textit{Range} \rangle \texttt{]})? \\
\langle \textit{CryptoLayer} \rangle &::= F \texttt{(} \langle \textit{Arg} \rangle (\texttt{,}\ \langle \textit{Arg} \rangle)^{*} \texttt{)} \\
\langle \textit{Arg} \rangle &::= (\langle \textit{Const} \rangle \mid \langle \textit{Var} \rangle \mid \langle \textit{FieldRef} \rangle \\
&\quad \mid f_{\text{aux}} \texttt{(} \langle \textit{Arg} \rangle^{+} \texttt{)} \mid \langle \textit{CalcExpr} \rangle) (\texttt{|}\ \langle \textit{Arg} \rangle)^{*}
\end{align*}

This structure contains four layers of semantics: the \textit{calculate} wrapping layer marks the execution boundary of cryptographic computation; the auxiliary function layer ($f_{\text{aux}}$) performs non-cryptographic operations such as truncation, encoding conversion, and concatenation; the cryptographic function layer ($F$) specifies a specific algorithm call, which can be an underlying primitive function ($F_{\text{prim}}$) or a composite function defined in a computation block ($F_{\text{def}}$); the argument reference layer traces each argument back to constants, variables, field references, nested auxiliary functions, or recursion.

\subsection{Session-Level Cryptographic Computation Grammar}

A single field's \textit{calculate} expression can represent local cryptographic computation, but session-level computations in cryptographic protocols often include multi-step sequential derivations. For example, TLS~1.3 needs to derive early\_secret, handshake\_secret, traffic\_secret, and finished\_key from the ECDHE shared secret \cite{rescorla2018transport}; IKEv2 needs to generate multiple key materials via prf+ \cite{harkins1998rfc2409}. To express such processes, CDSL introduces computation blocks, which are divided into flow-type and parameter-type.

Flow-type computation blocks are ordered composites of calculate expressions, arranging multiple assignment rules according to data dependencies into a sequentially executed rule chain. Its syntax is defined as follows:
\begin{align*}
\langle \textit{FlowBlock} \rangle &::= \textit{BlockName}\ \texttt{\{}\ \langle \textit{Rule} \rangle^{+}\ \texttt{\}} \\
\langle \textit{Rule} \rangle &::= \textit{Target}\ \texttt{=}\ \langle \textit{CalcExpr} \rangle \mid \langle \textit{DataExpr} \rangle \\
\langle \textit{DataExpr} \rangle &::= \langle \textit{AuxCall} \rangle \mid \langle \textit{ArgUnit} \rangle\ (\texttt{|}\ \langle \textit{ArgUnit} \rangle)^{*}
\end{align*}

Here, $\langle \textit{CalcExpr} \rangle$ follows the same four-layer nested grammar as transformational attributes $A_T$, and $\langle \textit{DataExpr} \rangle$ is a pure data expression (concatenation, slicing, length calculation, etc.) not involving cryptographic operations. Rules within the block are evaluated strictly in the written order; after the \textit{Target} of each rule is evaluated, it is injected into the context via $\Gamma$'s update operation, and subsequent rules can directly reference it via $\langle \textit{Var} \rangle$.

Parameter-type computation blocks describe algorithm configurations via declarative key-value pairs:
\begin{align*}
\langle \textit{ParamBlock} \rangle &::= \textit{BlockName}\ \texttt{\{}\ \langle \textit{Param} \rangle^{+}\ \texttt{\}} \\
\langle \textit{Param} \rangle &::= \textit{ParamKey}\ \texttt{:}\ (\langle \textit{Const} \rangle \mid \textit{null} \mid \textit{true} \\
&\quad \mid \textit{false})
\end{align*}

Each $\langle \textit{Param} \rangle$ corresponds to one configuration dimension for a key, used to declaratively define configuration parameters for operations such as signature algorithms.

\subsection{Comprehensive Example: TLS 1.3 Finished Message Construction}

\begin{figure}[htbp]
\centering
\includegraphics[scale=0.50, trim=10 10 10 10, clip]{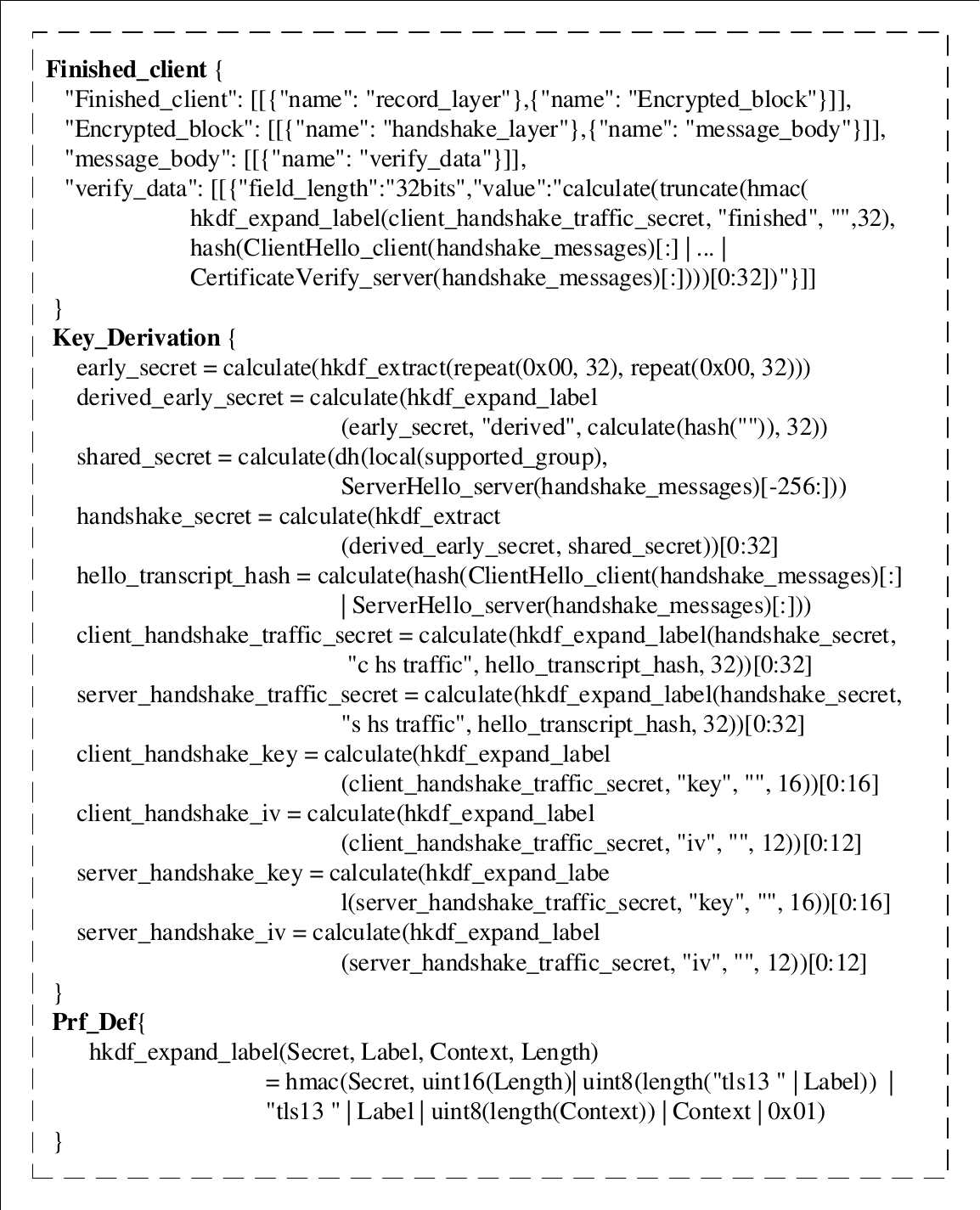}
\caption{The Finished\_client message part in PAST.}
\label{fig:Finished_client}
\end{figure}

Using the TLS~1.3 Finished\_client message as an example, we demonstrate how the three CDSL subsystems work together. Fig.~\ref{fig:Finished_client} shows the complete flow from CDSL description to message generation for the Finished message:

\begin{enumerate}
\item \replaced{SeriCrypt automatically generates the complete PAST under the constraints of CDSL, comprising the message structure tree $T$ carrying node attribute annotations and the set of session-level computation blocks $B$.}{The message structure grammar defines the nested topology of the Finished message, and the unified attribute evaluation system assigns attributes to each node;}

\item \replaced{The execution engine parses the nodes of the PAST. When it reaches the calculate expression of the verify\_data field in the Finished\_client message, a transformational attribute, it recognizes that the evaluation of verify\_data depends on handshake\_messages and client\_handshake\_traffic\_secret, and accordingly triggers the KeyDerivation and PrfDef computation blocks.}{The verify\_data field's calculate expression references the computation block variable client\_finished\_key and the handshake\_messages fields of each message in the runtime context.}

\item \replaced{KeyDerivation reads the DH key, ClientHello, ServerHello, and the preceding handshake messages from the runtime context $\Gamma$, sequentially derives keys such as early\_secret and handshake\_secret, and writes the results back to $\Gamma$.}{The session-level cryptographic computation grammar defines the KeyDerivation computation block and the PrfDef computation block, achieving explicit representation of client\_finished\_key.}

\item \added{The calculate expression of verify\_data reads the required key from $\Gamma$, invokes hkdf\_expand\_label, defined in the PrfDef computation block, to derive finished\_key, and computes verify\_data via HMAC.}

\item \added{The execution engine completes encryption encapsulation, length backfilling, and byte-level serialization, ultimately producing a Finished\_client message that conforms to the TLS~1.3 specification.}
\end{enumerate}

\added{Throughout this process, cross-message dependencies are resolved through attribute declarations and projections from the runtime context, and the computation blocks, once defined, are reused by the execution engine without protocol-specific code.}

\section{System Design and Implementation}
\label{sec:system}

\subsection{System Overview}
\label{sec:overview}

\begin{figure*}[htbp]
\centering
\includegraphics[scale=0.42, trim=10 10 10 10, clip]{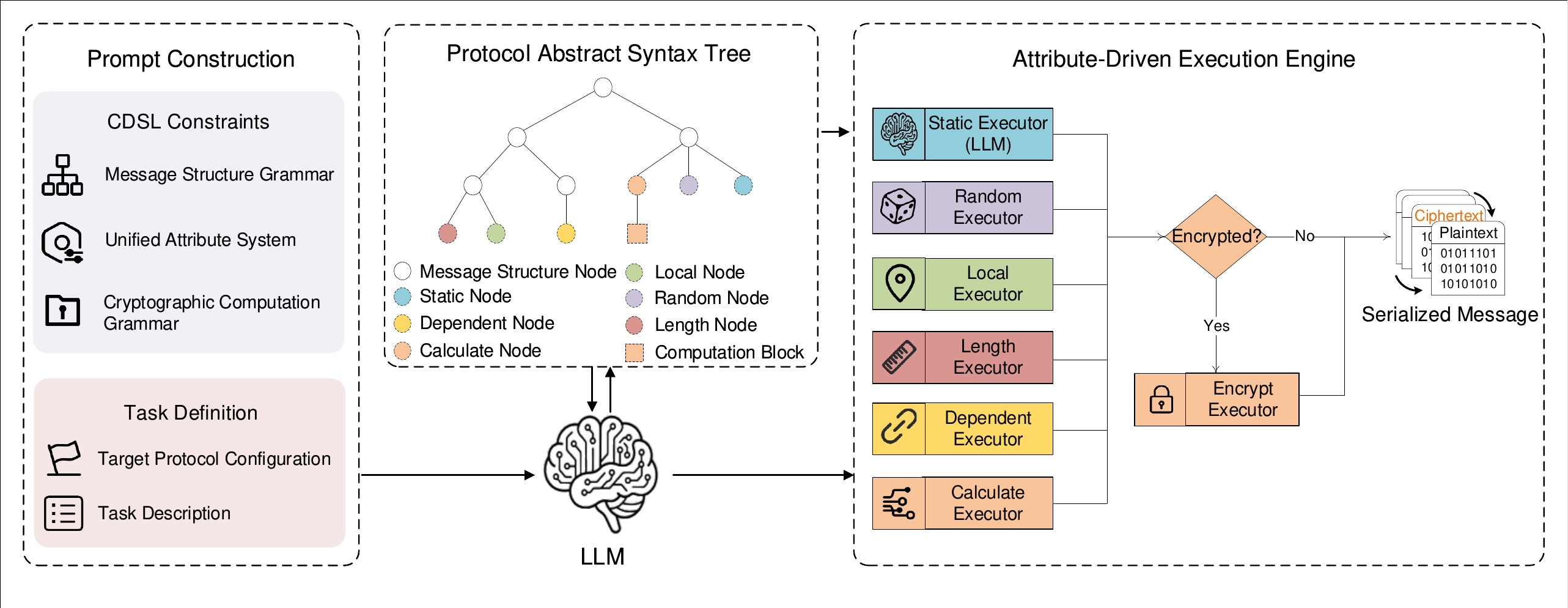}
\caption{SeriCrypt system operation flow.}
\label{fig:system-flow}
\end{figure*}

SeriCrypt divides the construction of cryptographic protocol messages into two phases: Phase1---LLM-driven PAST construction, and Phase2---attribute-driven PAST execution, with CDSL serving as the formal bridge connecting the two phases. \replaced{Moreover, the declarative structure of PAST supports protocol security testing: targeted mutations at the PAST level can construct specification-violating test inputs for violation detection (RQ5) and structured test cases for protocol fuzzing (RQ6).}{This two-phase architecture separates the functions of natural language understanding and deterministic computation: the LLM is responsible for the semantic parsing and structural induction of protocol specifications, thus avoiding the uncertainty in handling cryptographic computations; meanwhile, the execution engine remains protocol-independent, achieving universal support for different cryptographic protocols by parsing CDSL declarations. Moreover, the declarative structure of PAST provides support for protocol security testing---by applying targeted mutations to target messages at the PAST level, the system can construct test inputs that violate RFC security constraints, used for detecting the security of protocol implementations (see Section V-D).} Fig.~\ref{fig:system-flow} illustrates the full operation flow of the SeriCrypt system.

\subsection{Phase1: LLM-Driven PAST Construction}
\label{sec:phase1}

PAST construction adopts a chain-of-thought strategy~\cite{wei2022chain,zhou2022least}, decomposing the complex \replaced{specification}{RFC} understanding task into \replaced{three stages comprising eight sequential steps, each of which takes the output of the preceding step as input}{three progressive levels} (complete prompt templates are given in Appendix~B): 

\begin{enumerate}
\item\textbf {Message structure construction} --- \replaced{the LLM generates the packet composition of each client message following the protocol encapsulation hierarchy, and transforms it into grammar productions carrying field length constraints and initial attribute annotations;}{the LLM generates the complete packet composition according to the protocol specification and CDSL message structure grammar, expanded down to atomic fields following the protocol encapsulation hierarchy;}

\item\textbf {Attribute annotation} --- the LLM supplements each atomic field with attribute types and value rules according to the CDSL unified attribute evaluation system\added{, and through dependency tracing refines the value rules into references to concrete message fields}; 

\item\textbf {Cryptographic block definition extraction} --- the LLM extracts \replaced{the key derivation formulas, encryption flow, decryption flow, signature parameter configurations, and specialized PRF functions}{computation rules such as key derivation chains, encryption/decryption processes, pseudo-random functions, and signature parameter configurations} according to the protocol specification and CDSL session-level cryptographic computation grammar, and writes them into corresponding computation blocks.
\end{enumerate}

\added{To address errors that LLM hallucinations may introduce into the PAST, SeriCrypt sets up a verification loop: the generated messages are checked by the execution engine, and any PAST that fails the check is corrected by humans against the protocol specification field by field until the verification passes, thereby preventing errors in the extraction stage from propagating to downstream testing.}

\subsection{Phase2: Attribute-Driven PAST Execution}
\label{sec:phase2}

The execution engine is responsible for converting PAST into a sendable byte stream. The engine does not embed any protocol-specific message format or composite algorithm; its behavior is entirely driven by the structure rules, attribute declarations, and computation blocks in PAST, executed in the following three stages: 

\begin{enumerate}
\item\textbf {Tree instantiation and independent field evaluation} --- the engine expands the syntax tree instances in depth-first order according to the production set, solving static attributes (taken by LLM based on protocol knowledge), random attributes (generated by the underlying random number generator), and local attributes (value read from local files); 
\item\textbf {Dependent field and cryptographic computation evaluation} --- processes dependent (project value from $\Gamma$), length (sum subtree serialization length), and calculate (recursive expression engine evaluation); 
\item\textbf {Encryption encapsulation and serialization} --- loads the encryption computation block, replaces the plaintext subtree with ciphertext leaf nodes, then traverses depth-first, concatenating the byte values of all leaf nodes to output the final message.
\end{enumerate}

Protocol independence is a key design of the execution engine. The engine only embeds a set of atomic cryptographic primitives (hmac, dh, aes\_gcm\_encrypt, etc.) and does not provide protocol-specific composite algorithms. Composite functions in protocol specifications (such as TLS~1.3's HKDF-Expand-Label, IKEv2's prf+) are extracted by Phase1 as parameterized definitions and written into computation blocks. At runtime, the engine loads these computation blocks, expands the composite expressions into atomic primitives, and then uniformly evaluates them via the recursive expression engine. Protocol differences are thus encapsulated in the CDSL description layer, and no modification to the execution engine code is needed when extending to new protocols.


\deleted{RFC specifications use normative keywords such as MUST, MUST NOT, and SHALL NOT to identify security checks that implementations must perform~\cite{bradner1997key}. This section performs protocol security testing based on the declarative structure of PAST, detecting whether the target implementation correctly enforces these stipulations.}

\deleted{We systematically sorted out such security constraints from the specifications of the target protocols, and selected the subsets that could be tested via field mutations or construction of specific sequences. For field-level constraints, targeted modifications are applied at the PAST level: modifying the attribute values of atomic nodes to tamper with field values (e.g., replacing legacy\_version with a value not allowed by the specification), deleting specific subtrees to remove necessary protocol elements (e.g., deleting the key\_share extension), or modifying the input references of computation blocks to break cryptographic binding relationships (e.g., replacing the HMAC input of verify\_data with an incomplete handshake transcript digest). For sequence-level constraints, the tester can adjust the sending order of messages or skip specific interaction steps.}

\deleted{This design ensures that the mutated messages remain correct in encoding format and only violate constraints in the targeted semantic dimension---the characteristic of ``syntactically correct but semantically violating'' enables violation messages to pass the parsing stage of the server and reach the actual security check logic. During testing, the system executes all messages before the target message normally according to the original PAST to establish the session context, loads the mutated PAST at the target message and sends it, then observes the server's response. The judgment criterion is based on the normative requirements of the RFC: if the server correctly rejects, the implementation passes; if it does not reject as specified, it constitutes a potential security risk~\cite{barr2014oracle}.}

\section{Experimental Evaluation}

We evaluate SeriCrypt on \replaced{six cryptographic protocols across four
protocol families}{four cryptographic protocol versions across two protocol
families}. The evaluation revolves around the following research questions:

\begin{itemize}
\item \textbf{RQ1 \replaced{(Interoperability)}{(Correctness)}:} Can SeriCrypt automatically generate handshake messages that are accepted by real protocol implementations?
\item \textbf{RQ2 (Expressiveness):} Can CDSL's attribute system and computation block mechanism fully describe all message structures and cryptographic operations that the client needs to construct in the target protocols?
\item \textbf{RQ3 (LLM Generation Quality):} What is the accuracy of PAST generated by the LLM?
\item \textbf{RQ4 (LLM Comparison):} Can directly prompting the LLM to generate a complete protocol client successfully complete the handshake?
\item \textbf{RQ5 (Protocol Testing Application):} How practical is SeriCrypt in protocol testing?
\item \added{\textbf{RQ6 (Fuzzing Application):} How does SeriCrypt's cryptographic protocol serialization capability perform in protocol fuzzing scenarios?}
\end{itemize}

\subsection{Experimental Setup}

\textbf{Protocols and implementations.} The experiments cover \replaced{10}{8} protocol configurations: IKEv1 (PSK/certificate, Main Mode/Aggressive Mode), IKEv2 (PSK/certificate), TLS 1.2, \replaced{TLS 1.3, SSH (password authentication), and TLCP (dual-certificate mutual authentication)}{and TLS 1.3}, tested on \replaced{six}{four} mainstream open-source implementations: \replaced{strongSwan 5.9.14, Libreswan 4.12, OpenSSL 3.0.2, TLSe (commit e69790b), OpenSSH 9.8p1, and GmSSL 3.1.1}{strongSwan, Libreswan, OpenSSL, and TLSE}.

\deleted{\textbf{Baseline method.} We adopt ``direct LLM generation'' as the comparison baseline, i.e., asking the same LLM to directly generate runnable protocol client code. The LLM used is Gemini-3-Pro-Preview throughout.}

\replaced{\textbf{Verification method.} In protocol-testing scenarios, this work supports obtaining legitimate message sequences whose fields conform to the specification and that are accepted by real implementations as valid protocol inputs. To determine whether a generated message satisfies this requirement, we adopt two criteria: (1) the server accepts the message and returns the response prescribed by the normal protocol negotiation flow; and (2) Wireshark field-level parsing reveals no anomaly in the message.}{\textbf{Correctness judgment.} Handshake correctness is judged by two criteria: server acceptance --- the session completes without any Alert/Error response; Wireshark field-level parsing --- for all exchanged messages, field values are verified to be consistent with the protocol specification on a field-by-field basis.}

\replaced{\textbf{Auxiliary components.} Each protocol is equipped with a configuration file and a lightweight response parser of approximately 120 lines, the latter extracting the field values required for subsequent computations from the server's responses.}{\textbf{Auxiliary components.} Each protocol is equipped with a lightweight response parser (approximately 120 lines) to extract the field values required for subsequent computations from the server's return messages.}

\subsection{RQ1: \replaced{Interoperability}{Correctness}}

For each combination of protocol, server, and cipher suite, we run SeriCrypt to complete the full interaction.

\begin{table*}[htbp]
\centering
\caption{Handshake completion results.}
\label{tab:rq1}
\footnotesize
\begin{tabular}{cccccc}
\hline
Protocol & Auth Mode & Exchange Mode & Cipher Suite & Handshake & Wireshark \\
\hline
IKEv1 & PSK & Main Mode & aes128-sha1-modp1536 & \checkmark & \checkmark \\
IKEv1 & PSK & Aggressive Mode & aes128-sha1-modp1536 & \checkmark & \checkmark \\
IKEv1 & Certificate & Main Mode & aes128-sha1-modp1536 & \checkmark & \checkmark \\
IKEv1 & Certificate & Aggressive Mode & aes128-sha1-modp1536 & \checkmark & \checkmark \\
IKEv2 & PSK & --- & aes128gcm16-sha1-modp1536 & \checkmark & \checkmark \\
IKEv2 & Certificate & --- & aes128gcm16-sha1-modp1536 & \checkmark & \checkmark \\
TLS 1.2 & Certificate & --- & TLS\_DHE\_RSA\_WITH\_AES\_128\_GCM\_SHA256 & \checkmark & \checkmark \\
TLS 1.3 & Certificate & --- & TLS\_AES\_128\_GCM\_SHA256 & \checkmark & \checkmark \\
SSH & Password & --- & dh-group14-sha256-aes128-gcm & \checkmark & \checkmark \\
TLCP & Certificate & --- & ECDHE\_SM4\_GCM\_SM3 & \checkmark & \checkmark \\
\hline
\end{tabular}
\end{table*}

SeriCrypt successfully completed handshakes in all \replaced{10}{8} test configurations \added{(Table~\ref{tab:rq1})}, and all messages passed Wireshark field-level parsing. The server accepted the negotiation parameters, key materials, authentication fields, and encrypted payloads in the generated messages, demonstrating the \replaced{interoperability}{correctness} of SeriCrypt in automatically constructing cryptographic protocol messages.

\subsection{RQ2: Expressiveness}

We evaluate CDSL's expressiveness from two aspects: message structure and cryptographic operations. At the message structure level, CDSL described a total of \replaced{52}{41} client messages under \replaced{10}{8} protocol configurations, \replaced{with IKEv1, IKEv2, TLS 1.2, TLS 1.3, SSH, and TLCP accounting for 18, 12, 6, 5, 5, and 6 messages, respectively}{among which IKEv1 (PSK/certificate, Main/Aggressive Mode) accounted for 18 messages, IKEv2 (PSK/certificate) 12 messages, TLS 1.2 6 messages, and TLS 1.3 5 messages}. Every field could be mapped to one of CDSL's six attribute types, with no field pattern falling outside the attribute system's coverage. At the cryptographic operation level, the \replaced{six}{four} protocols involved a total of \replaced{28}{20} different cryptographic operations (see Table~\ref{tab:rq2} for details), all of which could be declared through CDSL computation blocks.

\begin{table*}[htbp]
\centering
\caption{Cryptographic operations covered by CDSL computation blocks.}
\label{tab:rq2}
\footnotesize
\begin{tabular}{cccccc}
\hline
Protocol & Key Derivation & Symmetric Encryption & Symmetric Decryption & Digital Signature & Pseudo-random Function \\
\hline
IKEv1 & SKEYID\_d/a/e & Enc\_IKEv1 & Dec\_IKEv1 & RSA & PRF \\
IKEv2 & SK\_d/ai/ar/ei/er/pi/pr & Enc\_IKEv2 & Dec\_IKEv2 & RSA-PKCS1v15 & PRF+ \\
TLS 1.2 & Write\_MAC\_Key, Write\_Key/IV & Enc\_TLS1.2 & Dec\_TLS1.2 & RSA-PKCS1v15 & TLS1.2\_PRF (P\_hash) \\
TLS 1.3 & Handshake\_Key/IV & Enc\_TLS1.3 & Dec\_TLS1.3 & RSA-PSS & Hkdf\_Expand\_Label \\
SSH & Enc\_Key/IV & Enc\_SSH & Dec\_SSH & -- & -- \\
TLCP & Write\_Key/IV & Enc\_TLCP & Dec\_TLCP & SM2 & TLCP\_PRF (P\_SM3) \\
\hline
\end{tabular}
\end{table*}

\deleted{CDSL's six attribute types covered all client message field patterns, and the computation block mechanism covered all 20 categories of cryptographic operations involved in the target protocols, indicating that CDSL has sufficient expressiveness for cryptographic protocol serialization tasks.}

\subsection{RQ3: LLM Generation Accuracy for PAST}

For each target message in the \replaced{six protocols}{four protocols}, we used the Phase1 pipeline with \replaced{two LLMs, Gemini-3-Pro-Preview and ChatGPT-5.5, each of which repeatedly generated PAST five times. The accuracy of each stage was measured on the generated PAST before correction: during the manual verification described in Section V-B, every generated PAST was reviewed against the protocol specification, the errors found in the LLM output were recorded, and the accuracy was computed based on the correctly generated items.}{Gemini-3-Pro-Preview to repeatedly generate PAST five times and recorded the average accuracy of each stage. Structure generation accuracy was determined by field-by-field comparison with the RFC; attribute annotation and computation block extraction accuracy were independently reviewed by four master's students majoring in cybersecurity, with disagreements jointly adjudicated by the first author and the four students.}

\begin{table}[htbp]
\centering
\caption{PAST generation accuracy in each stage.}
\label{tab:rq3}
\footnotesize
\setlength{\tabcolsep}{3.5pt}
\begin{tabular}{ccccc}
\hline
Protocol & LLM & Structure & Attribute & Computation Block \\
\hline
IKEv1   & Gemini-3-Pro-Preview & 98.6\% & 94.9\% & 93.1\% \\
IKEv1   & ChatGPT-5.5          & 94.2\% & 92.8\% & 92.4\% \\
IKEv2   & Gemini-3-Pro-Preview & 95.5\% & 93.9\% & 95.8\% \\
IKEv2   & ChatGPT-5.5          & 94.1\% & 91.3\% & 92.5\% \\
TLS 1.2 & Gemini-3-Pro-Preview & 98.7\% & 94.8\% & 95.8\% \\
TLS 1.2 & ChatGPT-5.5          & 97.4\% & 95.1\% & 97.9\% \\
TLS 1.3 & Gemini-3-Pro-Preview & 92.1\% & 93.3\% & 98.0\% \\
TLS 1.3 & ChatGPT-5.5          & 93.4\% & 93.4\% & 95.5\% \\
SSH     & Gemini-3-Pro-Preview & 96.7\% & 96.7\% & 92.9\% \\
SSH     & ChatGPT-5.5          & 95.1\% & 95.0\% & 91.6\% \\
TLCP    & Gemini-3-Pro-Preview & 96.1\% & 98.0\% & 94.7\% \\
TLCP    & ChatGPT-5.5          & 96.2\% & 96.2\% & 94.4\% \\
\hline
\end{tabular}
\end{table}

\replaced{Both LLMs exhibit similar error characteristics. As shown in Table~\ref{tab:rq3}, the structure generation accuracy of both LLMs is the lowest for TLS 1.3, because TLS 1.3 moves many functions into ClientHello extensions, which are numerous, deeply nested, and subject to complex optional conditions, so the LLM easily misjudges which extensions must appear. The attribute annotation accuracy of IKEv1/v2 and TLS 1.3 is relatively low, as their field values are scattered across specific payloads of multiple exchanges, making value tracing difficult. The computation block extraction accuracy of SSH is the lowest, because its key derivation adopts a character-labeled hash chain, whose operand order and splitting details are error-prone. We analyzed the errors recorded during the manual verification of both LLMs across the six protocols, and the errors of both models fall into the same three categories. \textbf{E1 (misjudgment of conditional field presence):} optional fields are numerous, and the prompts cannot precisely cover every optional part, so their presence is easily misjudged. \textbf{E2 (cross-message state and transcript boundary):} when a field value depends on complex message context, dependency tracing sometimes cannot reach the exact source location. \textbf{E3 (cryptographic derivation detail error):} the overall structure of the derivation is correct, but the operand order, trailing bytes, or splitting layout is wrong. Overall, although the above errors commonly occur in LLM-generated PAST, the average accuracy remains above 94\%, and these errors can all be discovered and corrected through the verification loop described in Section V-B, demonstrating the value of the LLM in the SeriCrypt framework.}{The structure generation accuracy for TLS 1.3 was the lowest, at 89.1\%. The main reason is that TLS 1.3 moves a large number of functions into ClientHello extension fields. There are many extension fields, deep nesting, and complex optional conditions. The LLM easily misjudges which extensions must appear under a given configuration and also easily misses field boundaries in deep variable-length structures. The computation block extraction accuracy for IKEv1 was the lowest, at 88.0\%. The reason comes from IKEv1's complex IV generation mechanism. Among Main Mode, Aggressive Mode, and Quick Mode, the IV calculation methods for the first encrypted message and subsequent encrypted messages in each mode are not the same; the LLM easily mixes up or misses IV rules from different phases. Overall, the LLM can effectively generate PAST. The average accuracy of the structure generation and attribute annotation stages exceeds 95\%, and the average accuracy of computation block extraction is about 94.5\%. Errors are mainly concentrated in protocol locations with highly complex structures or lengthy key derivation chains, and these errors can be discovered and corrected during the verification stage through field comparison and computation rule review.}

\subsection{RQ4: Comparison with Direct LLM Generation}

\replaced{We adopt ``direct LLM generation'' as the comparison baseline, i.e., asking the same LLM to directly generate runnable protocol client code; the LLM used for this baseline is Gemini-3-Pro-Preview. To test this baseline, we designed a progressive prompting strategy to generate a complete protocol client for each target protocol.}{We designed a progressive prompting strategy to test the LLM's ability to directly generate a complete protocol client.} The prompting process decomposes the task into steps such as message structure definition, structure construction and serialization, key derivation, encrypted payload construction, and interaction flow control. Each target protocol was repeated 5 times, for a total of \replaced{30}{20} sets of experiments. The generated results were classified into five levels (L0 --- code cannot run, L1 --- message rejected, L2 --- partial plaintext negotiation completed, L3 --- encryption construction failed, L4 --- full handshake successful).

\begin{table}[htbp]
\centering
\caption{Direct LLM generation results.}
\label{tab:rq4}
\footnotesize
\begin{tabular}{ccc}
\hline
Protocol & Best Level & Main Failure Cause \\
\hline
IKEv1 & L2 & Key derivation failure \\
IKEv2 & L2 & Negotiation material acquisition failure \\
TLS 1.2 & L3 & Encrypted payload construction failure \\
TLS 1.3 & L2 & Key derivation failure \\
SSH & L2 & Key derivation failure \\
TLCP & L2 & Key derivation failure \\
\hline
\end{tabular}
\end{table}

In all \replaced{30}{20} sets of experiments, none reached L4 (full handshake successful) \added{(Table~\ref{tab:rq4})}. We categorized the observed failures into three types: (1) Protocol constant or encoding convention errors --- the LLM can identify the general structure of message types and fields but easily miswrites constant values, byte order, and length encoding. (2) Historical message negotiation value acquisition failure --- the code generated by the LLM often fails to stably maintain cross-message state, potentially ignoring the cipher suite selected by the server or omitting random numbers. (3) Cryptographic computation chain breakage --- key derivation and encryption encapsulation are the primary failure sources; the LLM easily confuses HKDF label concatenation format, PRF input order, and the scope of AEAD additional authenticated data.

These results corroborate SeriCrypt's division-of-labor strategy. \replaced{The LLM excels at extracting protocol knowledge from protocol specifications (RQ3 indicates that the average accuracy of PAST generation exceeds 94\%). CDSL's design precisely separates specification extraction from computation execution: the LLM handles specification understanding and knowledge extraction, which it is relatively good at; the execution engine handles precise computation and byte-level encoding, which the LLM struggles to perform stably.}{The LLM excels at extracting message structures and attribute semantics from RFC documents (RQ3 indicates that the average accuracy of structure generation and attribute annotation exceeds 95\%), but cannot reliably implement the full byte-level details of cryptographic computation chains. CDSL's design precisely separates these two layers: the LLM handles specification understanding and structure extraction, which it is relatively good at; the execution engine handles precise computation and byte-level encoding, which the LLM struggles to perform stably.} This conclusion echoes existing work~\cite{li2025cipherbank,fakhoury20253dgen,meng2024large,sun2026semfuzz}; the experimental results further show that when the task boundary extends from format extraction to cryptographic computation, the LLM's reliability rapidly declines. The combination of declarative intermediate representation and deterministic execution engine can mitigate this capability gap.

\subsection{RQ5: Protocol Testing Application}

\replaced{We extracted security constraints identified by normative keywords such as MUST, MUST NOT, SHOULD, and SHOULD NOT~\cite{bradner1997key} (and their equivalents in the TLCP national standard~\cite{gbt38636}) from the specifications of the six target protocols, selected the subset that could be violated via field-level PAST mutations, and guided the LLM to construct multiple test cases for each constraint. During testing, the system first executes all messages before the target message normally according to the original PAST to establish the session context, and then loads and sends the mutated PAST at the target message. Mutations are expressed at the declarative PAST level while encoding is still handled by the execution engine, so the mutated messages remain syntactically correct but violate the targeted constraints, thus reaching the server's actual security check logic; if the server does not reject as the specification requires, a potential violation is flagged~\cite{barr2014oracle}.}{We extracted security constraints identified by normative keywords such as MUST, MUST NOT, SHOULD, SHOULD NOT from the RFC specifications of the four target protocols, selected the subset that could be violated via PAST field-level mutations, and guided the LLM to construct multiple test cases violating each constraint.}

\begin{table}[htbp]
\centering
\caption{Security constraint testing results.}
\label{tab:rq5}
\footnotesize
\begin{tabular}{cccc}
\hline
Protocol & Security Constraints & Test Cases & Security Violations \\
\hline
IKEv1 & 27 & 66 & 0 \\
IKEv2 & 26 & 79 & 1 \\
TLS 1.2 & 20 & 54 & 0 \\
TLS 1.3 & 16 & 61 & 4 \\
SSH & 19 & 55 & 0 \\
TLCP & 17 & 49 & 0 \\
\hline
Total & 125 & 364 & 5 \\
\hline
\end{tabular}
\end{table}

Based on the \replaced{125}{89} security constraints extracted from \replaced{the specifications}{the RFCs}, \replaced{364}{260} test cases were constructed, and 5 security violations were discovered \added{(Table~\ref{tab:rq5})}, of which 4 were new findings \deleted{and have been reported to the developers}. A detailed analysis of the security violations is provided in Appendix~C. The results indicate that PAST's declarative structure can support targeted testing of protocol security constraints, verifying the practicality of SeriCrypt in protocol testing scenarios.

\begin{table*}[htbp]
\centering
\caption{Summary of security violations.}
\label{tab:violations}
\footnotesize
\begin{tabular}{cccc}
\hline
\textbf{ID} & \textbf{Implementation (Protocol)} & \textbf{Violation} & \textbf{Security Impact} \\
\hline
V-1 & strongSwan (IKEv2) & No response to INFORMATIONAL request & Session maintenance silently interrupted \\
V-2 & TLSE (TLS 1.3) & Use of non-negotiated signature algorithm & Negotiation constraint broken, risking weak algorithm selection \\
V-3 & TLSE (TLS 1.3) & Failure to reject MD5-signed certificate & Weak-hash certificate passes authentication \\
V-4 & TLSE (TLS 1.3) & Failure to check required extension & Handshake proceeds without a required extension \\
V-5 & TLSE (TLS 1.3) & Acceptance of illegal ChangeCipherSpec value & Malformed message silently ignored \\
\hline
\end{tabular}
\end{table*}

\added{The five violations (Table~\ref{tab:violations}) are concentrated in certificate verification, extension checking, and session maintenance, which indicates that implementations tend to relax the enforcement of RFC security constraints at these checkpoints; V-1 was disclosed in prior work, while V-2 through V-5 are new findings of this work and have been reported to the developers.}

\added{\subsection{RQ6: Fuzzing Application}}

\added{To evaluate SeriCrypt's performance in protocol fuzzing scenarios, we augment SeriCrypt with a basic structured mutation component, which performs structured mutations on PAST fields to generate test cases. We compare SeriCrypt with the mainstream fuzzing tools AFLnet~\cite{pham2020aflnet} and ChatAFL~\cite{meng2024large} on six cryptographic protocols --- IKEv1, IKEv2, TLS 1.2, TLS 1.3, SSH, and TLCP --- and measure the code coverage achieved within the same time budget of four hours.}

\begin{table}[htbp]
\centering
\caption{Code coverage (AFL bitmap \%) achieved under 4-hour fuzzing.}
\label{tab:rq6}
\footnotesize
\begin{tabular}{cccc}
\hline
Protocol (Implementation) & SeriCrypt & AFLnet & ChatAFL \\
\hline
IKEv2 (strongSwan)         & 29.38\% & 23.14\% & 22.85\% \\
IKEv1 (strongSwan)         & 27.51\% & 19.05\% & 19.01\% \\
TLS 1.3 (OpenSSL)          & 24.87\% & 23.81\% & 23.83\% \\
TLS 1.2 (OpenSSL)          & 25.31\% & 21.72\% & 21.69\% \\
SSH (OpenSSH)              & 8.19\% & 6.62\% & 6.60\% \\
TLCP (GmSSL)               & 12.05\% & 9.11\% & 9.02\% \\
\hline
\end{tabular}
\end{table}

\added{SeriCrypt achieves the highest code coverage on all six protocols (Table~\ref{tab:rq6}). These protocols impose strict requirements on the cryptographic correctness of messages, and most blind byte-level mutations fail to pass the server's cryptographic checks, whereas PAST-based structured mutations preserve cryptographic validity, enabling the test cases to reach deeper protocol states. The results indicate that SeriCrypt's cryptographic protocol serialization capability effectively supports protocol fuzzing, and that the time overhead of dynamically parsing declarations and evaluating the attribute tree does not impact overall throughput, as SeriCrypt achieves the highest coverage within the same time budget.}

\section{Discussion}

\subsection{\replaced{Why CDSL}{Relationship with Format Description Languages}}
\replaced{The structural description layer of CDSL overlaps with typed JSON schemas and with format description languages such as 3D~\cite{ramananandro2019everparse} and Kaitai~\cite{kaitai2026documentation}, all of which declaratively express field types, length constraints, and nesting levels, and differs from them only in where the declarative boundary is drawn. JSON Schema draws this boundary at the logical-structure level: it cannot represent binary encoding, so the binary encoding, field dependencies, and cryptographic semantics required for cryptographic protocol modeling are all undertaken by accompanying code; Kaitai and 3D push the boundary further, to binary layout and intra-message computed fields. Yet what lies beyond the boundary, cross-message session state and cryptographic semantics, is precisely the core of cryptographic protocols, and no modest extension of the intra-message expressive model of existing languages can cover these two requirements, which can only be realized in code. CDSL pushes the declarative boundary further still, to cross-message state and cryptographic semantics, and through reusable computation blocks decomposes protocol-specific composite algorithms into atomic cryptographic primitives, achieving protocol independence on the basis of the representation of cryptographic protocols.}{The structural description layer of CDSL has overlap with format description languages such as 3D~\cite{ramananandro2019everparse} and Kaitai~\cite{kaitai2026documentation} — they all express field types, length constraints, and nesting levels in a declarative manner. However, the main boundary of these languages stops at message parsing or static format description; cryptographic semantics are implemented by developers on a per-protocol basis. CDSL incorporates cryptographic semantics into the declarative description scope through its attribute type system and computation block mechanism, enabling the execution engine to construct complete chains without writing protocol-specific code.}

\subsection{Extensibility}

\added{The favorable extensibility to new protocols is a significant advantage of SeriCrypt over existing methods such as Scapy~\cite{biondi2024scapy} and TLS-Attacker~\cite{somorovsky2016systematic}.} The current evaluation \replaced{covers the TLS, IKE, SSH, and TLCP protocol families}{covers the TLS and IKE protocol families} which differ significantly in key derivation, message encapsulation, and exchange mechanisms. The results of RQ2 show that CDSL's six attribute types and computation block mechanism have sufficient expressiveness for these differences. When extending SeriCrypt to new protocols, the core logic of the execution engine is fully reused, \replaced{and the manual effort consists of three parts: (1) constructing a configuration file that specifies session parameters such as the addresses of both endpoints, the selected algorithm types, and the message types to be generated, together with an output example of 5--8 fields; (2) implementing a lightweight response parser of approximately 120 lines; (3) manually verifying and correcting the LLM-generated PAST.}{developers only need to provide lightweight response parsers.}

\subsection{Future Directions}

\replaced{SeriCrypt currently focuses on the serialization of client messages. Automatically generating response parsers based on the CDSL framework is the next focus of work. Meanwhile, we plan to further validate the extensibility of SeriCrypt on a broader range of cryptographic protocols (e.g., QUIC~\cite{mandl2024quic}, WireGuard~\cite{donenfeld2017wireguard}, and post-quantum hybrid handshakes) and diverse cryptographic algorithm combinations. These protocols often have more complex specifications or less coverage in the training data, so the error types summarized in RQ3 may become more pronounced, and the verification loop described in Section V-B will play a more important role.}{SeriCrypt currently focuses on the serialization of client messages. Automatically generating response parsers based on the CDSL framework is the next focus of work. Meanwhile, we plan to further validate the extensibility of SeriCrypt on a broader range of cryptographic protocols (e.g., SSH~\cite{ylonen2006secure}, QUIC~\cite{mandl2024quic}) and diverse cryptographic algorithm combinations.} Furthermore, combining PAST mutation strategies with structure-aware fuzzing~\cite{sun2026semfuzz,aschermann2019nautilus,srivastava2021gramatron} is a natural evolution direction\added{, and the preliminary results of RQ6 support this direction. Another promising direction is to replace the manual verification of LLM-generated PAST with agents (e.g., Claude Code) that iteratively test and repair the PAST, further reducing the human effort.}

\section{Related Work}

Table~\ref{tab:comparison} compares the capability boundaries of SeriCrypt with representative approaches across four dimensions: D1 (automated specification knowledge extraction), D2 (cross-message dependency expression), D3 (cryptographic computation expression), and D4 (protocol-independent execution engine). ChatAFL~\cite{meng2024large} uses LLMs to extract protocol state machines from RFCs to guide fuzzing. 3DGen~\cite{fakhoury20253dgen} combines LLMs with format description languages to generate verified binary format parsers. ParCleanse~\cite{zheng2025validating} extracts DSL-format constraints from RFCs and combines them with the Z3 solver to generate constraint-satisfying packets. SemFuzz~\cite{sun2026semfuzz} extracts constraints from RFCs to implement structured mutations on plaintext messages. These works demonstrate that LLMs can effectively participate in extracting protocol format knowledge and state machine knowledge, but their downstream tasks do not involve cross-message cryptographic state maintenance. TLS-Attacker~\cite{somorovsky2016systematic} and Scapy~\cite{biondi2024scapy} provide byte-level message construction capabilities, but the cryptographic logic is written in imperative code, and supporting each new protocol or cipher suite requires writing substantial protocol-specific code, making cross-protocol reuse difficult\replaced{;}{.}\added{in protocol security testing, constructing semantic-level mutations such as malformed messages that involve cross-message dependencies or cryptographic computations likewise requires locating and modifying the corresponding computation logic layer by layer, at a substantially higher implementation cost than SeriCrypt. We quantitatively verify this difference with a controlled experiment (see Appendix~D for details).}

\begin{table}[htbp]
\centering
\caption{Comparison between SeriCrypt and existing methods.}
\label{tab:comparison}
\begin{tabular}{ccccc}
\hline
\textbf{Method} & \textbf{D1} & \textbf{D2} & \textbf{D3} & \textbf{D4} \\
\hline
TLS-Attacker  & $\times$ & $\circ$ & $\circ$ & $\times$ \\
Scapy         & $\times$ & $\circ$ & $\circ$ & $\times$ \\
ChatAFL       & $\checkmark$ & $\circ$ & $\times$ & -- \\
3DGen         & $\checkmark$ & $\times$ & $\times$ & $\checkmark$ \\
ParCleanse    & $\checkmark$ & $\times$ & $\times$ & $\checkmark$ \\
SemFuzz       & $\checkmark$ & $\times$ & $\times$ & $\checkmark$ \\
SeriCrypt     & $\checkmark$ & $\checkmark$ & $\checkmark$ & $\checkmark$ \\
\hline
\end{tabular}

\medskip
\noindent
\textit{Note:} $\checkmark$ = support, $\times$ = no support, $\circ$ = partial support, -- = not applicable.
\end{table}

\section{Conclusion}

This paper proposes SeriCrypt, a context-aware serialization framework for network cryptographic protocols. SeriCrypt designs a domain-specific language for cryptographic protocols, CDSL, which uniformly describes message structures, field attributes, cross-message context references, and cryptographic computation rules. It proposes an LLM-based specification extraction process that converts \replaced{unstructured specification text}{unstructured RFC text} into PAST. It implements a protocol-independent general execution engine that, driven by attribute declarations, completes field evaluation and message serialization. Experimental results show that SeriCrypt can generate valid messages accepted by real protocol implementations, completing handshakes in all \replaced{10}{8} test configurations. Directly prompting the LLM to generate client code failed to achieve a single complete handshake. The average accuracy of LLM-generated PAST exceeds \replaced{94}{94.5}\%. Based on PAST, targeted violation tests against security constraints successfully constructed test cases and discovered 5 security violations, verifying the framework's practicality in protocol security testing scenarios. \added{In protocol fuzzing, SeriCrypt can generate test sequences that pass the server's cryptographic checks, enabling fuzzing to reach deeper protocol states.} The above results indicate that the division-of-labor strategy of ``LLM extracts structure, deterministic engine executes computation'' is applicable to scenarios such as cryptographic protocols that impose strict requirements on state consistency and computational correctness. The cryptographic protocol message construction capability provided by SeriCrypt can further serve protocol security testing tasks such as fuzzing, state machine learning, and conformance checking.

\section*{Acknowledgment}

We would like to thank our shepherd and the anonymous reviewers for their thoughtful feedback. This work is supported by the National Key Research and Development Program of China under Grant 2023YFA1009500.

\bibliographystyle{IEEEtran}
\bibliography{ref}

\newpage
\section*{APPENDIX}

\subsection*{Appendix A}
\subsubsection*{Example of Message Structure Grammar}
\begin{figure}[htbp]
\centering
\includegraphics[scale=0.44, trim=10 10 10 10, clip]{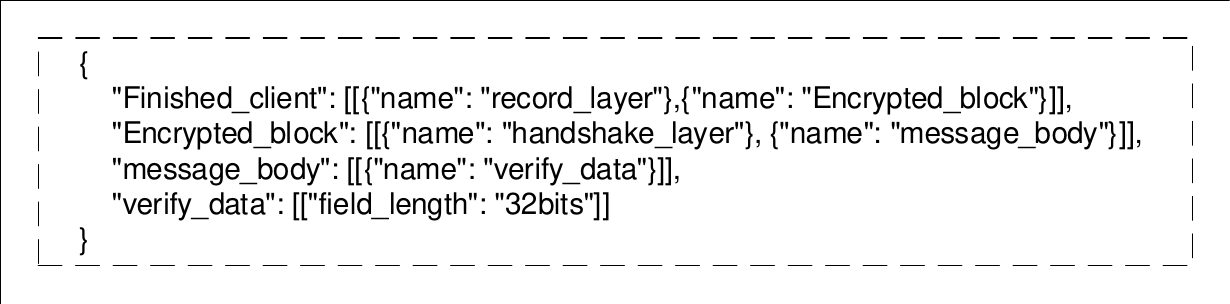}
\caption{TLS 1.3 Finished\_Client message structure fragment.}
\label{fig:structure fragment}
\end{figure}

As shown in Fig.~\ref{fig:structure fragment}, the message structure of TLS 1.3 Finished\_Client is illustrated in detail, including field composition and field lengths.
\subsubsection*{Example of Unified Attribute Evaluation System}
\begin{figure}[htbp]
\centering
\includegraphics[scale=0.44, trim=10 10 10 10, clip]{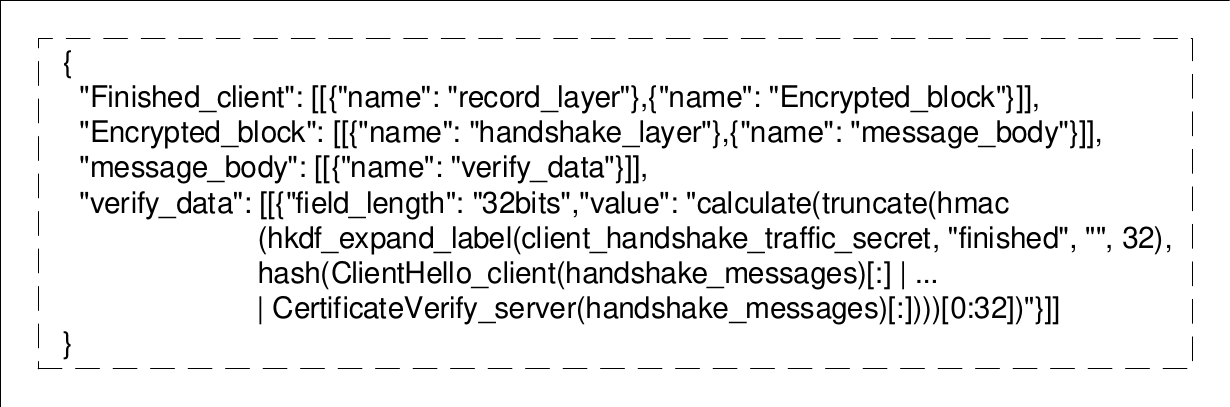}
\caption{Attributes of fields in TLS 1.3 Finished\_Client.}
\label{fig:Attributes}
\end{figure}

As shown in Fig.~\ref{fig:Attributes}, the attribute annotation results of UAS on the message fields of TLS 1.3 Finished\_Client are presented.
\subsubsection*{Example of Session-Level Cryptographic Computation Grammar}
\begin{figure}[htbp]
\centering
\includegraphics[scale=0.44, trim=10 10 10 10, clip]{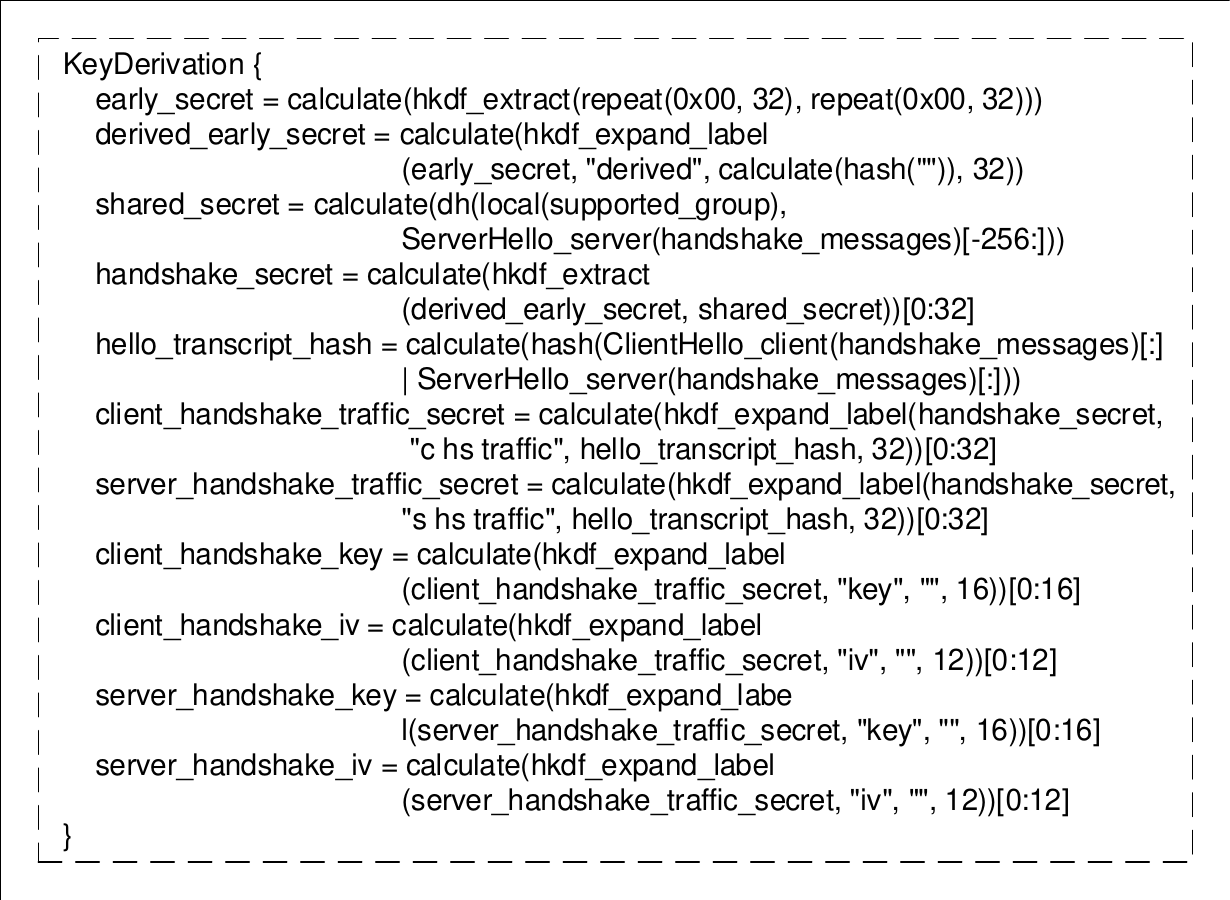}
\caption{TLS 1.3 Handshake key derivation.}
\label{fig:Key Derivation}
\end{figure}

\begin{figure}[H]
\centering
\includegraphics[scale=0.44, trim=10 10 10 10, clip]{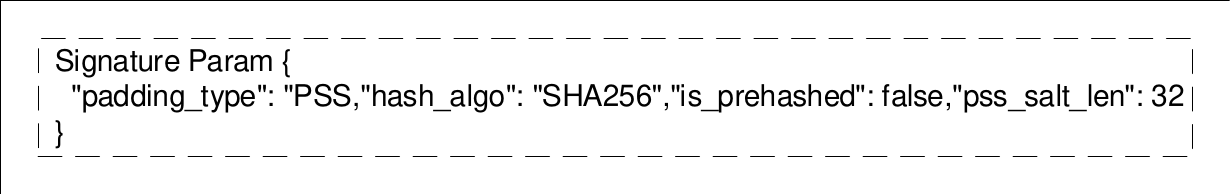}
\caption{TLS 1.3 Signature algorithm block.}
\label{fig:Signature Algorithm}
\end{figure}

Fig.~\ref{fig:Key Derivation} and Fig.~\ref{fig:Signature Algorithm}, illustrate the process-oriented and parameter-oriented computation blocks, describing the TLS 1.3 key derivation process and signature algorithm parameters, respectively.
\subsection*{Appendix B}
\begin{figure}[htbp]
\centering
\includegraphics[scale=0.44, trim=10 10 10 10, clip]{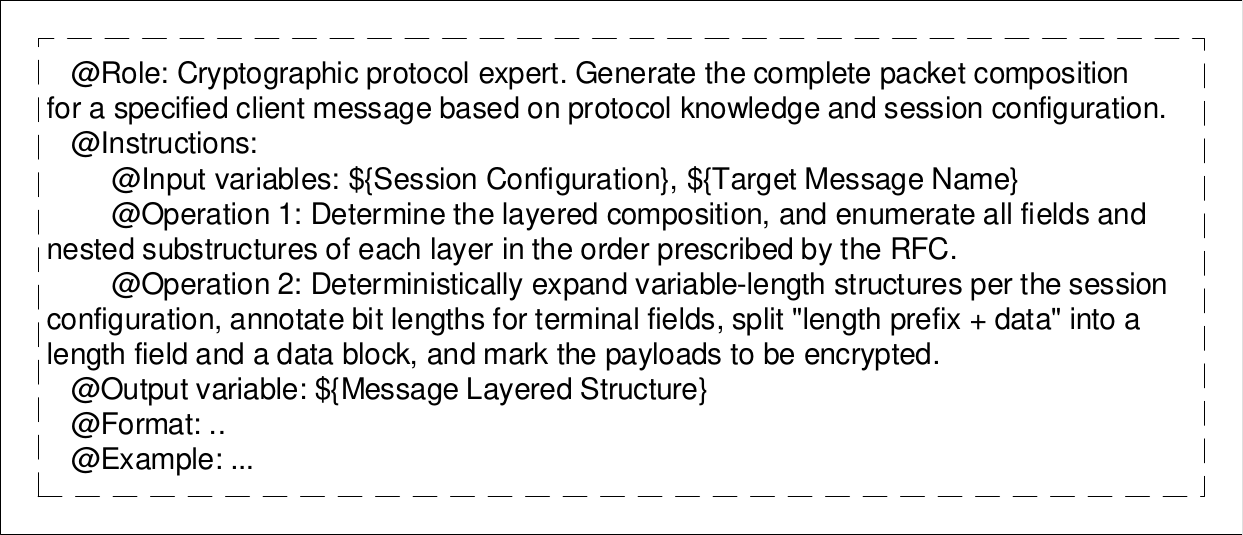}
\caption{Packet composition generation prompt template.}
\label{fig:packet composition}
\end{figure}

\begin{figure}[htbp]
\centering
\includegraphics[scale=0.44, trim=10 10 10 10, clip]{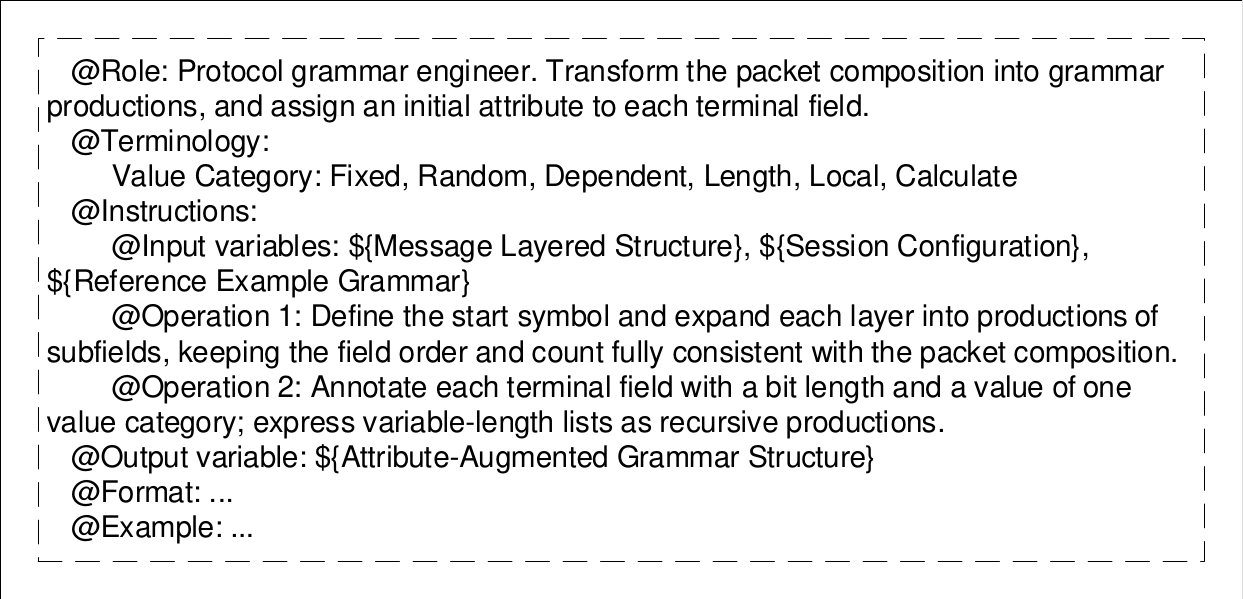}
\caption{Grammar structure generation prompt template.}
\label{fig:grammar generation}
\end{figure}

\begin{figure}[htbp]
\centering
\includegraphics[scale=0.44, trim=10 10 10 10, clip]{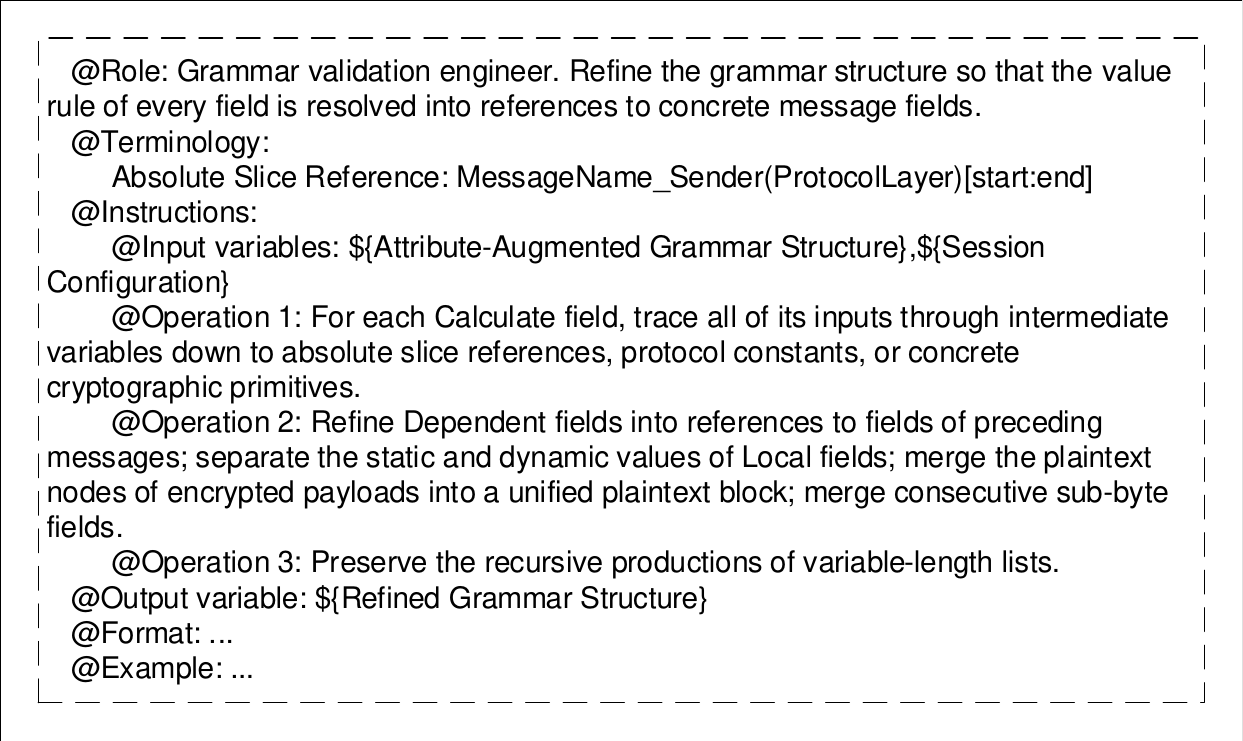}
\caption{Attribute refinement prompt template.}
\label{fig:attribute refinement}
\end{figure}

\begin{figure}[htbp]
\centering
\includegraphics[scale=0.44, trim=10 10 10 10, clip]{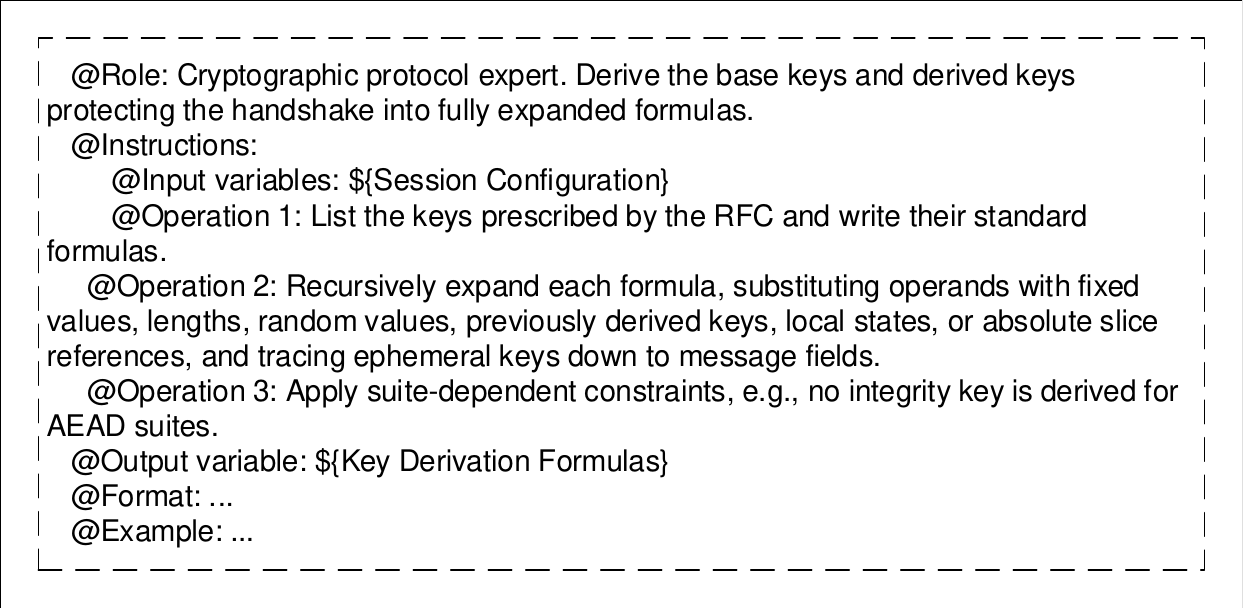}
\caption{Key derivation formula prompt template.}
\label{fig:key derivation}
\end{figure}

\begin{figure}[htbp]
\centering
\includegraphics[scale=0.44, trim=10 10 10 10, clip]{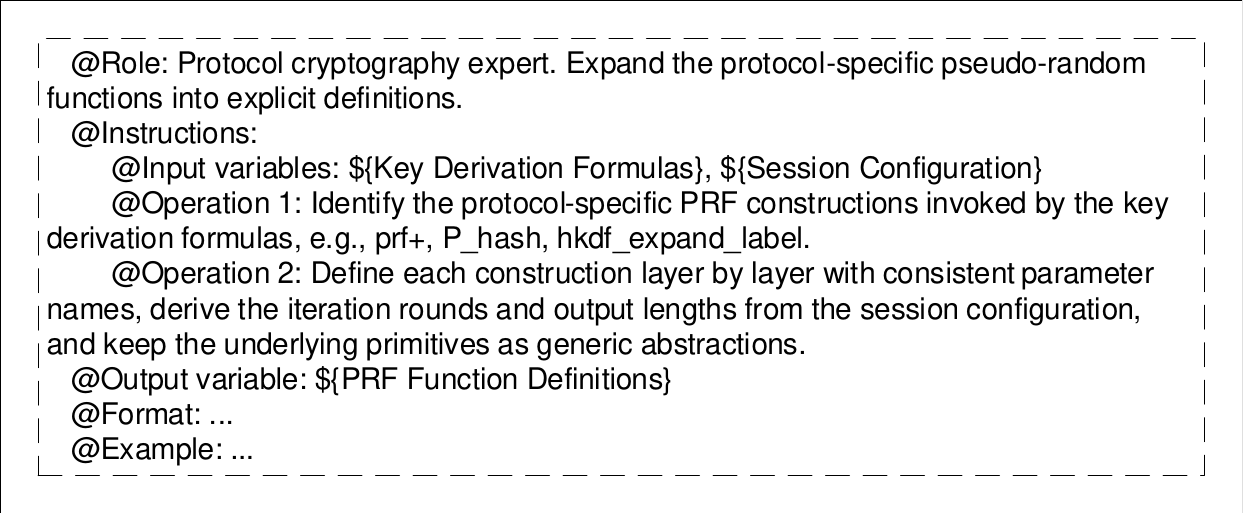}
\caption{Encryption flow prompt template.}
\label{fig:encryption flow}
\end{figure}

\begin{figure}[htbp]
\centering
\includegraphics[scale=0.44, trim=10 10 10 10, clip]{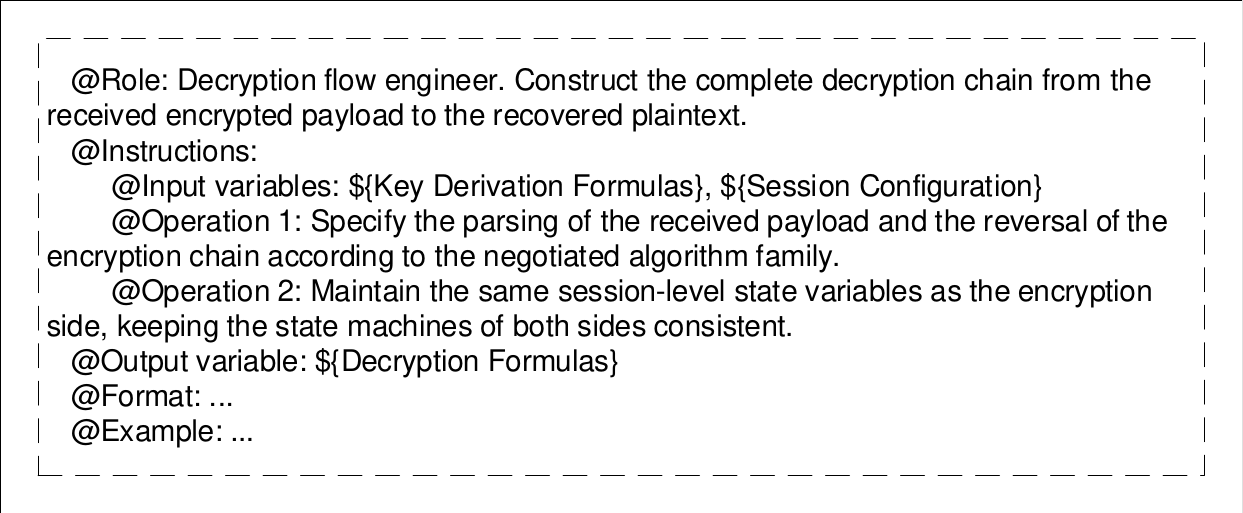}
\caption{Decryption flow prompt template.}
\label{fig:decryption flow}
\end{figure}

\begin{figure}[htbp]
\centering
\includegraphics[scale=0.44, trim=10 10 10 10, clip]{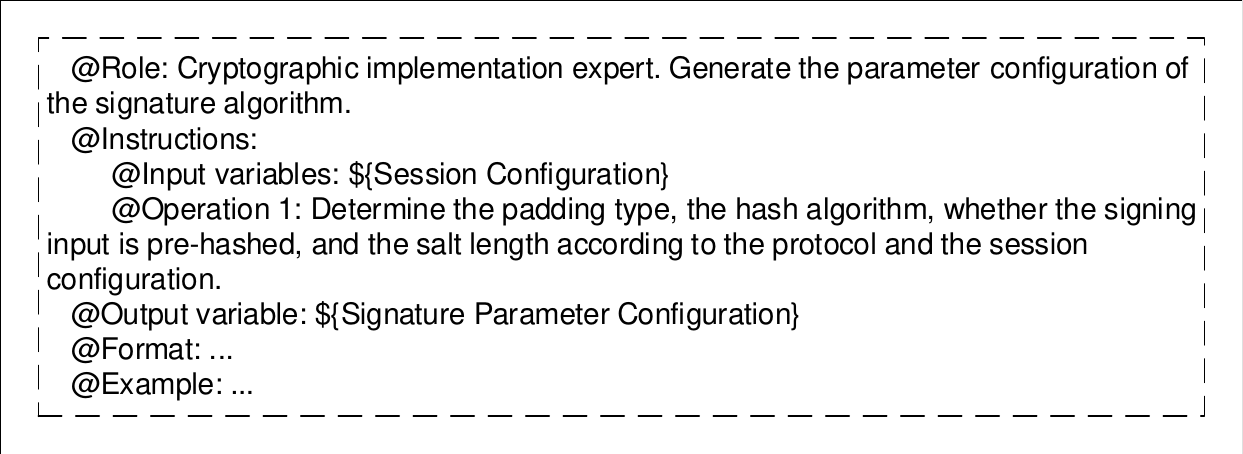}
\caption{Signature parameter configuration prompt template.}
\label{fig:signature configuration}
\end{figure}

\begin{figure}[htbp]
\centering
\includegraphics[scale=0.44, trim=10 10 10 10, clip]{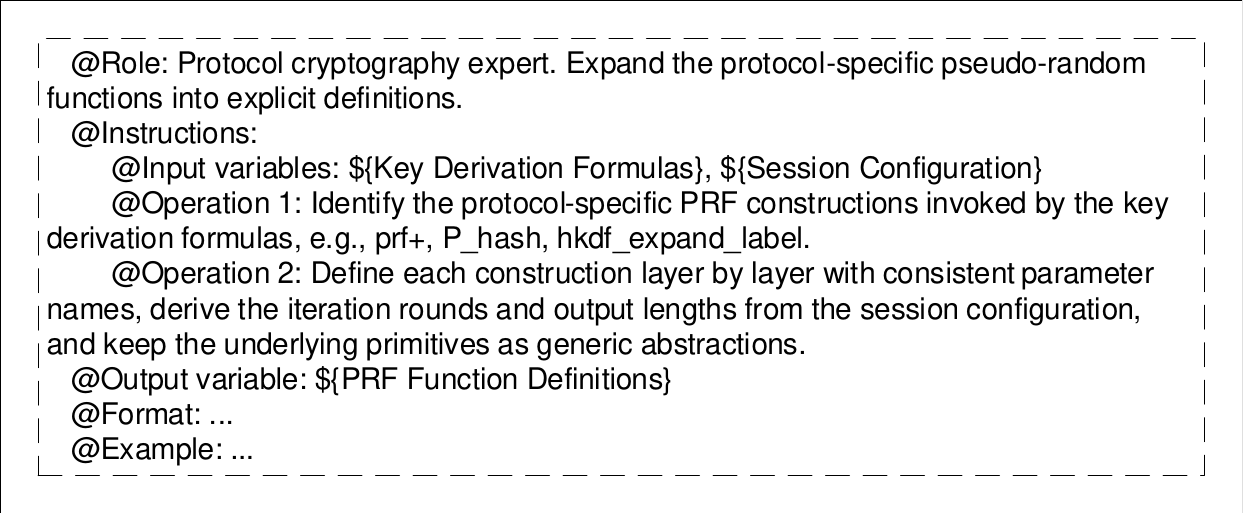}
\caption{PRF specialization function prompt template.}
\label{fig:prf specialization}
\end{figure}

\replaced{The prompt templates of the eight steps are presented in detail in Fig.~\ref{fig:packet composition} through Fig.~\ref{fig:prf specialization}, following the execution order, where Figs.~\ref{fig:packet composition} and~\ref{fig:grammar generation} correspond to message structure construction, Fig.~\ref{fig:attribute refinement} corresponds to attribute annotation, and Figs.~\ref{fig:key derivation} through~\ref{fig:prf specialization} correspond to computation block definition extraction.}{As shown in Fig.~\ref{fig:Message structure construction}, Fig.~\ref{fig:Attribute annotation} and Fig.~\ref{fig:Computation block definition}, the prompt templates for the three processes --- message structure construction, attribute annotation, and computation block definition --- are presented in detail.}

\subsection*{Appendix C}

This appendix summarizes the 5 semantic violations detected by SeriCrypt during the experiments. \replaced{The RFC basis, test construction method, and observation result of each violation are detailed below.}{First, Table~\ref{tab:violations} provides an overview of the violations; subsequently, the RFC basis, test construction method, and observation results for each violation are explained item by item.}

\textbf{V-1 No Response to INFORMATIONAL Request}

RFC Specification: RFC~7296 Section~1: ``The types of subsequent exchanges are CREATE\_CHILD\_SA (which creates a Child SA) and INFORMATIONAL (which deletes an SA, reports error conditions, or does other housekeeping). Every request requires a response.''

Test Construction: SeriCrypt first establishes a normal IKEv2 session, then sends a malformed encrypted payload. After strongSwan returns an error and updates its internal state, SeriCrypt continues to send an INFORMATIONAL request.

Observed Result: strongSwan did not return any response to the INFORMATIONAL request. According to RFC~7296, the receiver MUST return a response to every INFORMATIONAL request.

\textbf{V-2 Server Selected a Signature Algorithm Not Offered by the Client}

RFC Specification: RFC~8446 Section~4.4.3: ``If the CertificateVerify message is sent by a server, the signature algorithm MUST be one offered in the client's `signature\_algorithms' extension unless no valid certificate chain can be produced without unsupported algorithms (see Section~4.2.3).''

Test Construction: SeriCrypt constructs a TLS~1.3 ClientHello message whose 
signature\_algorithms extension does not include the signature algorithm subsequently used by the server.

Observed Result: TLSE still selected a signature algorithm not declared as supported by the client during the handshake and did not abort the handshake.

\textbf{V-3 Acceptance of MD5-Signed Certificate}

RFC Specification: RFC~8446 Section~4.4.2.2: ``Any endpoint receiving any certificate which it would need to validate using any signature algorithm using an MD5 hash MUST abort the handshake with a `bad\_certificate' alert.''

Test Construction: SeriCrypt presents a certificate signed with the MD5 algorithm to TLSE.

Observed Result: TLSE did not send a bad\_certificate alert to abort the handshake.

\begin{table*}[t]
\centering
\caption{Implementation cost of the nine mutation tasks. Each cell reports
LoC / number of functions.}
\label{tab:mutation_cost}
\begin{tabular}{ccccc}
\hline
Task & Target deviation & SeriCrypt & Scapy & TLS-Attacker \\
\hline
S1 & client\_version = 0x0304 & 1 / 0 & 7 / 0 & 1 / 1 \\
S2 & cipher\_suites = [0x00FF] & 1 / 0 & 7 / 0 & 1 / 1 \\
S3 & Finished record length under-counted by 4 & 1 / 0 & 12 / 1 & 5 / 1 \\
D1 & Server Certificate excluded from the transcript hash of verify\_data & 1 / 0 & 13 / 1 & 19 / 2 \\
D2 & client\_random excluded from the master\_secret seed & 1 / 0 & 6 / 1 & 1 / 1 \\
D3 & ClientKeyExchange excluded from the CertificateVerify signature transcript & 1 / 0 & 13 / 1 & 11 / 1 \\
C1 & Finished PRF label swapped & 1 / 0 & 14 / 1 & 1 / 1 \\
C2 & CertificateVerify signature SHA256$\to$SHA1 & 2 / 0 & 8 / 1 & 1 / 1 \\
C3 & GCM tag replaced by the first 16 ciphertext bytes & 1 / 0 & 8 / 1 & 1 / 1 \\
\hline
Total & & 10 / 0 & 88 / 7 & 41 / 10 \\
\hline
\end{tabular}
\end{table*}

\textbf{V-4 Acceptance of ClientHello Missing Required Extension}

RFC Specification: RFC~8446 Section~9.2: ``If not containing a `pre\_shared\_key' extension, it MUST contain both a `signature\_algorithms' extension and a `supported\_groups' extension. \ldots Servers receiving a ClientHello which does not conform to these requirements MUST abort the handshake with a `missing\_extension' alert.''

Test Construction: SeriCrypt constructs a non-PSK mode TLS~1.3 ClientHello and removes the signature\_algorithms extension, retaining only the supported\_groups extension.

Observed Result: TLSE did not send a missing\_extension alert, and the handshake continued.

\textbf{V-5 Tolerance of Illegal ChangeCipherSpec}

RFC Specification: RFC~8446 Section~5.1: ``An implementation which receives any other change\_cipher\_spec value or which receives a protected change\_cipher\_spec record MUST abort the handshake with an `unexpected\_message' alert.''

Test Construction: SeriCrypt constructs an illegal ChangeCipherSpec message, modifying its payload value from the legal value 0x01 to the illegal value 0x02.

Observed Result: TLSE did not send an unexpected\_message alert, but instead ignored the malformed CCS message and continued the handshake.

\subsection*{\added{Appendix D}}

\added{\textbf{Experimental Environment.} TLS 1.2, cipher suite TLS\_DHE\_RSA\_WITH\_AES\_128\_GCM\_SHA256, mutual authentication, with OpenSSL s\_server as the server under test.}

\added{\textbf{Task Set.} Nine tasks covering three categories---structural fields (S1--S3), cross-message dependencies (D1--D3), and cryptographic computations (C1--C3); the target deviation of each task is identical across the three tools (Table~\ref{tab:mutation_cost}).}

\added{\textbf{Metrics.} The lines of code (LoC) written or modified and the number of functions defined or modified to implement each mutation; the cost of building the baseline client is excluded.}

\added{All three tools completed all mutations and generated the corresponding malformed binary messages. The difference lies in the implementation cost: SeriCrypt requires a total of 10 lines and 0 function changes---each task is accomplished by editing 1--2 declaration lines---amounting to 1/4.1 and 1/8.8 of the totals of TLS-Attacker and Scapy, respectively; the difference is largest for dependency-affecting tasks such as D1/D3 (3 lines versus 32 and 31), where imperative tools must rewrite the parsing of the handshake transcript and the hash recomputation logic, whereas SeriCrypt only needs to modify a single concatenation term in the declaration expression.}

\end{document}